\documentclass[aps, prd, onecolumn, tightenlines, notitlepage, superscriptaddress, nofootinbib, preprintnumbers, floatfix,showkeys,11pt]{revtex4-2}

\usepackage[normalem]{ulem}
\usepackage{amstext}
\usepackage{amssymb}
\usepackage{amsmath}
\usepackage{graphicx}
\graphicspath{{plots/}}
\usepackage{url}
\usepackage{color}
\usepackage{ulem}
\usepackage[utf8]{inputenc}
\pdfoutput=1
\usepackage{textcomp}
\usepackage{comment}
\usepackage{yfonts}
\usepackage{epsfig,amsfonts,mathrsfs,amsmath,amssymb,graphicx,color,slashed,multirow}
\usepackage{amsmath,latexsym,amssymb,graphicx,slashed,color,enumerate,url,cancel,gensymb}
\usepackage{textcomp}

\usepackage[x11names]{xcolor}
\usepackage[colorlinks]{hyperref}

\usepackage{booktabs}
\usepackage{adjustbox}

\definecolor{vdrgreen}{rgb}{0.0, 0.7, 0.0}

\definecolor{indianred}{rgb}{0.8, 0.36, 0.36}
\definecolor{blue(ncs)}{rgb}{0.0, 0.53, 0.74}
\AtBeginDocument{\hypersetup{citecolor=indianred,linkcolor=indianred,urlcolor=indianred}}

\usepackage{float}

\usepackage[T1]{fontenc}
\usepackage{ae,aecompl}
\graphicspath{{Figures/}}
\usepackage{appendix}



\newcommand{\AddrIISERB}{Department of Physics, Indian Institute of Science Education and Research - Bhopal, \\ 
Bhopal Bypass Road, Bhauri, Bhopal 462066, India}

\newcommand{\AddrAthens}{%
Department of Physics, National and Kapodistrian University of Athens, Zografou Campus GR-15772 Athens, Greece}

\newcommand{\AddrAHEP}{%
 Instituto de F\'{i}sica Corpuscular (CSIC-Universitat de Val\`{e}ncia), Parc Cient\'ific UV C/ Catedr\'atico Jos\'e Beltr\'an, 2 E-46980 Paterna (Valencia) - Spain}

\usepackage{orcidlink}
\begin{document}

\title{\LARGE XENONnT and LUX-ZEPLIN constraints \\ on  DSNB-boosted dark matter}
\author{Valentina De Romeri}\email{deromeri@ific.uv.es}
\affiliation{\AddrAHEP}
\author{Anirban Majumdar}\email{anirban19@iiserb.ac.in}
\affiliation{\AddrIISERB}
\author{Dimitrios K. Papoulias}\email{dkpapoulias@phys.uoa.gr}
\affiliation{\AddrAthens}
\author{Rahul Srivastava}\email{rahul@iiserb.ac.in}
\affiliation{\AddrIISERB}


\begin{abstract}
We consider a scenario in which dark matter particles are accelerated to semi-relativistic velocities through their scattering with the Diffuse Supernova Neutrino Background. Such a subdominant, but more energetic dark matter component can be then detected via its scattering on the electrons and nucleons inside direct detection experiments. This opens up the possibility to probe the sub-GeV mass range, a region of parameter space that is usually not accessible at such facilities. We analyze current data from the XENONnT and LUX-ZEPLIN experiments and we obtain novel constraints on the scattering cross sections of sub-GeV boosted dark matter with both nucleons and electrons. We also highlight the importance of carefully taking into account Earth's attenuation effects as well as the finite nuclear size into the analysis. By comparing our results to other existing constraints, we show that these effects lead to improved and more robust constraints. 

\end{abstract}
\maketitle

\section{Introduction}
\label{sec:intro}

It is estimated that $85$\% of the matter content of the Universe is in the form of a hypothetical kind of matter, dubbed dark matter (DM)~\cite{Planck:2018vyg}. One of the biggest mysteries in contemporary physics and astronomy is to understand its microscopic nature. 
However, since DM does not interact with photons and interacts very ``weakly'' with ordinary matter, it proves challenging to detect it. On the other hand, DM gravitational effects on visible matter
allow us to infer its existence despite its elusiveness.  
One of the most compelling solutions to the DM puzzle assumes it to be in the form of some unknown particle~\cite{Bertone:2004pz}, thus calling for an extension of the Standard Model (SM). Several strategies, including direct and indirect detection experiments and collider searches, have been developed to try to detect it~\cite{Queiroz:2016awc, Gninenko:2023pkv}.
Although a conclusive finding of DM has not been achieved yet, these searches have imposed very tight constraints on its potential properties. As part of the continuous effort to understand this enigmatic component, new experiments, and observations are being carried out.

The possibility that DM has been produced thermally in the early Universe and that its abundance is determined by thermal freeze-out has motivated numerous large direct detection (DD) experiments, which aim at observing the scattering of a DM particle off a target in a deep underground detector. These experiments have experienced an increasingly, decades-long progress which has brought them into the multi-ton era~ \cite{Billard:2021uyg}. Current most-sensitive constraints in the high-mass regime include those set by liquid xenon (LXe) experiments like LUX-ZEPLIN (LZ)~\cite{LZ:2023poo,LUX-ZEPLIN:2022xrq}, XENONnT~\cite{XENON:2022ltv,XENONCollaboration:2023orw}, XENON1T~\cite{XENON:2018voc}, PandaX-II~\cite{PandaX-II:2020oim} and LUX~\cite{LUX:2016ggv}, together with measurements on liquid argon (LAr) detectors like DEAP-360~\cite{DEAP:2019yzn} and DarkSide-50~\cite{DarkSide:2018kuk} and the solid-state cryogenic detector of SuperCDMS~\cite{SuperCDMS:2017mbc}.

We are interested in the results recently released by two DD experiments, XENONnT~\cite{XENON:2022ltv} and LZ~\cite{LUX-ZEPLIN:2022xrq}. Both experiments use state-of-the-art LXe detectors that aim at observing low-energy electron and nuclear recoils induced by DM scattering.
Being one of the most sensitive DM DD experiments at present,
XENONnT~\cite{XENON:2015gkh,XENON:2020kmp}, installed at the Gran Sasso National Laboratories in Italy, is the upgrade phase of XENON1T~\cite{XENON:2018voc}. 
Thanks to its larger active target mass, superior photon detection mechanism, and extremely low background, XENONnT is an order of magnitude more sensitive to weakly-interacting DM particles than its predecessor.
The recently released XENONnT data 
correspond to a total exposure of 1.16 tonne$\times$years~\cite{XENON:2022ltv}.
The LZ experiment~\cite{LZ:2019sgr},
located at the Sanford Underground Research Facility in South Dakota, is a detector centered on a dual-phase time projection chamber, also filled with LXe.  
 The recently available LZ data 
 correspond to an exposure of 5.5 tonne$\times$60 days~\cite{LUX-ZEPLIN:2022xrq}. The LZ collaboration has reported results from a blind search for DM particles and established the current strongest constraint for masses above
9 GeV, testing a cross section as small as $6\times 10^{-48}\mathrm{~cm}^2$ at a DM mass of $30$ GeV.

Both XENONnT and LZ, as most of other DD experiments, have best sensitivities to electroweak-scale DM with masses around 10-100 GeV. Below the GeV scale, their sensitivity drops dramatically, as the electron and nuclear recoil energy becomes smaller and eventually falls below the detector threshold. 
Normally, recoil events in the LZ experiment cannot be observed for non-relativistic sub-GeV DM traveling at velocities $v \sim 10^{-3}$. However, an energetic sub-GeV DM particle may generate a substantial signal. 
One possibility that has been put forward to explore sub-GeV DM is that of boosted DM (BDM). Such a BDM would contribute as a subdominant component of the total DM flux, but would nonetheless enhance the mass reach of DD experiments, allowing to explore the sub-GeV range.
This idea has first been proposed considering DM boosting from the scattering with energetic galactic cosmic rays~\cite{Bringmann:2018cvk,Ema:2018bih} and has been extensively discussed in the literature, see for instance~\cite{Cappiello:2018hsu,Dent:2019krz,Alvey:2019zaa, Lei:2020mii,Dent:2020syp,Xia:2021vbz, CDEX:2022fig,Alvey:2022pad,Arguelles:2022fqq,Maity:2022exk,Cappiello:2019qsw,Jho:2020sku,Cho:2020mnc, PandaX-II:2021kai, Su:2023zgr, Herrera:2023nww}. More recently, the possibility that DM is boosted through its scattering with neutrinos has also been envisaged, either considering cosmic neutrinos~\cite{Jho:2021rmn}, solar neutrinos~\cite{Zhang:2020nis}, neutrinos from primordial black holes evaporation~\cite{Chao:2021orr,Calabrese:2021zfq,Calabrese:2022rfa,Li:2022jxo}, supernova neutrinos~\cite{Lin:2022dbl,Lin:2023nsm} or the diffuse supernova neutrino background (DSNB)~\cite{Das:2021lcr,Ghosh:2021vkt}. Other possibilities leading to BDM include blazar-boosted DM~\cite{Bhowmick:2022zkj}, boosted DM from phantom dark energy~\cite{Cline:2023hfw}, confined cosmic rays in starburst galaxies~\cite{Ambrosone:2022mvk}, models with semi-annihilating DM~\cite{DEramo:2010keq,Berger:2014sqa, Toma:2021vlw} or models with a multi-component DM sector~\cite{Agashe:2014yua, Kong:2014mia, Aoki:2018gjf, Borah:2021yek}. 

In this work, we investigate the possibility that the DM in the Milky Way halo is boosted to semi-relativistic velocities, via its scattering on the DSNB~\cite{Beacom:2010kk,Lunardini:2010ab}. The DSNB is a cumulative and isotropic flux of MeV neutrinos of all flavors produced from core-collapse supernovae explosions along the whole history of the Universe. While not yet observed, the DSNB is an irreducible background, expected to be within the reach of near-future experiments. Even though less energetic than cosmic rays, it seems reasonable to assume possible interactions of local DM with this isotropic neutrino background. By employing XENONnT and LZ latest data releases~\cite{XENON:2022ltv, LUX-ZEPLIN:2022xrq}, we derive stringent constraints on both DM-electron and DM-nucleon scattering cross sections in the sub-GeV range, thus providing complementary results to the standard analyses offered by the two collaborations~\cite{XENONCollaboration:2023orw, LUX-ZEPLIN:2022xrq}. We highlight and pay special attention to the Earth's attenuation effects, that, as we will show, play an important role in the region of interest of the parameter space. Additionally, we also take into account nuclear effects which further improve the sensitivity and robustness of our analysis. DSNB-boosted DM had previously been considered in Ref.~\cite{Das:2021lcr} as a possible explanation to an excess of electron recoil events in the low energy
region, now disappeared, observed by XENON1T~\cite{XENON:2020rca}. Ref.~\cite{Ghosh:2021vkt} also set limits on DSNB-boosted DM scattering off electrons using XENON1T and Super-Kamiokande data. Here we improve upon these previous results by presenting for the first time constraints on DSNB-boosted DM, from the most recent XENONnT and LZ data, for both nuclear and electron scattering. 

The remainder of this paper is organized as follows. Section~\ref{Sec:DSNB_Flux} provides a discussion on theoretical predictions for the DSNB flux. Sec.~\ref{Sec:Boosted_DM} explains how sub-GeV non-relativistic DM particles in the Milky Way halo can attain semi-relativistic speeds due to interactions with DSNB neutrinos, and highlights the importance of Earth's attenuation effects as well as the nuclear form factors. In Sec.~\ref{Sec:DM_Scattering}, we delve into the simulation of the DSNB-boosted DM-induced signal predicted for  the XENONnT and LZ detectors. Our results are presented in Sec.~\ref{sec:results}, while we finally provide our concluding remarks in Sec.~\ref{sec:conclusions}.
%

\section{\label{Sec:DSNB_Flux} Theoretical estimate of the DSNB flux}

Right after the first star formation event, the Universe has been surrounded by an isotropic flux of MeV-energy neutrinos and antineutrinos of all flavors, produced from all supernovae events from the core-collapse explosions of huge stars throughout the Universe.  
The theoretical prediction for the differential DSNB flux, per neutrino flavor $\alpha$, 
can be estimated as~\cite{Horiuchi:2008jz, Beacom:2010kk,DeGouvea:2020ang,Ando:2023fcc}
\begin{equation}
\label{equn:DSNB_diff_Flux}
\frac{d\Phi_{\nu_\alpha}^\mathrm{DSNB}}{dE_\nu}=\int_0^{\text{z}_\mathrm{max}}d\text{z}\frac{R_\mathrm{CCSN}(\text{z})}{H(\text{z})}\left.\mathscr{F}_{\nu_\alpha}(E_{\nu_\alpha})\right|_{E_\nu=E^s_\nu(1+\text{z})} \,,
\end{equation}
$E^s_\nu$ being the neutrino energy at the source. The integral is performed over the redshift parameter, $\text{z}$, and we take the maximum redshift at which star-formation occurs as $\text{z}_\mathrm{max}\sim 6$. Moreover, $H(\text{z})$ is the Hubble function determined from the Friedmann equation as
\begin{equation}
\label{equn:Hubble_Constant}
H(\text{z})=H_0\sqrt{\Omega_M(1+\text{z})^3+\Omega_\Lambda(1+\text{z})^{3(1+w)}+ (1 - \Omega_M - \Omega_\Lambda)(1+\text{z})^2} \,,
\end{equation}
where $H_0=67.45~\mathrm{km~s}^{-1}\mathrm{Mpc}^{-1}$ is the Hubble constant~\cite{Planck:2018vyg,Workman:2022ynf}, $\Omega_M=0.315 \pm 0.007$ and $\Omega_\Lambda=0.685 \pm 0.007$  denote the matter and vacuum  contributions to the present-Universe energy density, 
while the best current
measurement for the equation-of-state parameter for the dark energy is $w=-1.028 \pm 0.031$~\cite{Planck:2018vyg}. The DSNB flux further depends upon the rate of Core-Collapse Supernovae (CCSN), which reads~\cite{DeGouvea:2020ang}
\begin{equation}
\label{equn:RCCSN}
R_\mathrm{CCSN}(\text{z})=\dot{\rho}_*(\text{z}) \frac{\int_8^{50}\psi(M)dM}{\int_{0.1}^{100}M\psi(M)dM}\,,
\end{equation}
where $\psi(M)$ is the initial mass function (IMF) of stars, indicating the star density within a certain mass range. For our analysis we have assumed the IMF to be a power-law distribution, $\psi(M)\propto M^{-2.35}$ according to~\cite{Salpeter:1955it}. The redshift evolution of the co-moving cosmic star-formation rate, $\dot{\rho}_*(\text{z})$, can be modelled as~\cite{Horiuchi:2008jz, Yuksel:2008cu}
\begin{equation}
\label{equn:SFR}
\dot{\rho}_*(\text{z})=\dot{\rho}_0\left[(1+\text{z})^{-10 a}+\Big(\frac{1+\text{z}}{B}\Big)^{-10 b}+\Big(\frac{1+\text{z}}{C}\Big)^{-10 c}\right]^{-0.1}\,,
\end{equation}
where the overall normalization factor is $\dot{\rho}_0=0.0178^{+0.0035}_{-0.0036}~ \mathrm{M}_\odot\mathrm{yr}^{-1}\mathrm{Mpc}^{-3}$~\cite{DeGouvea:2020ang}. The constants $B$, and $C$ are expressed as~\cite{Horiuchi:2008jz, Yuksel:2008cu}:
\begin{subequations}
    \begin{equation}
        B=(1+\mathrm{z}_1)^{1-\frac{a}{b}}\,,
    \end{equation}
    \begin{equation}
        C=(1+\mathrm{z}_1)^{\frac{b-a}{c}} (1+\mathrm{z}_2)^{1-\frac{b}{c}}\,,
    \end{equation}
\end{subequations}
where $\mathrm{z}_1=1$, and $\mathrm{z}_2=4$ represent the redshift breaks, while $a$, $b$ and $c$ denote the logarithmic slopes for the low, intermediate, and high redshift ranges. An analytical fit to data from different astronomical surveys~\cite{Yuksel:2008cu, Horiuchi:2008jz} gives 
$$\{a, b, c \} = \{3.4 \pm 0.2, -0.3 \pm 0.2,-3.5 \pm 1 \} \, .$$  Finally, a non-degenerate Fermi-Dirac distribution is used to parametrize the flavor-dependent neutrino spectra released by a CCSN event~\cite{Beacom:2010kk,DeGouvea:2020ang}
\begin{equation}
\label{equn:FD_Distribution}
\mathscr{F}_{\nu_\alpha}(E_{\nu_\alpha})=\frac{E_\nu^\mathrm{tot}}{6}\frac{120}{7\pi^4}\frac{E_{\nu_\alpha}^2}{T_{\nu_\alpha}^4}\frac{1}{1+e^{E_{\nu_\alpha}/T_{\nu_\alpha}}}\,,
\end{equation}
where $E_\nu^\mathrm{tot}=3\times 10^{53}\mathrm{erg}$,\footnote{It is assumed that the total released energy, $E_\nu^\mathrm{tot}$, is equally distributed among the 6 neutrino flavors, since during CCSN the neutrino emission mostly occurs in the cooling phase, which persists for about $10$ s  after the bounce.} represents the total amount of energy released as neutrinos~\cite{Beacom:2010kk}, and $T_{\nu_\alpha}$ denotes the temperature of each flavor of neutrinos. In our present study, we consider $T_{\nu_e}=6.6\mathrm{~MeV,~} T_{\bar{\nu}_e}=7\mathrm{~MeV,~} T_{\nu_x}=10\mathrm{~MeV}$ ($\nu_x$ denotes either $\nu_\mu$ or $\nu_\tau$ or their antiparticles), satisfying the upper limit extracted from Super-Kamiokande~\cite{Super-Kamiokande:2013ufi}.

\begin{figure}[t]
 \includegraphics[width=0.65 \textwidth]{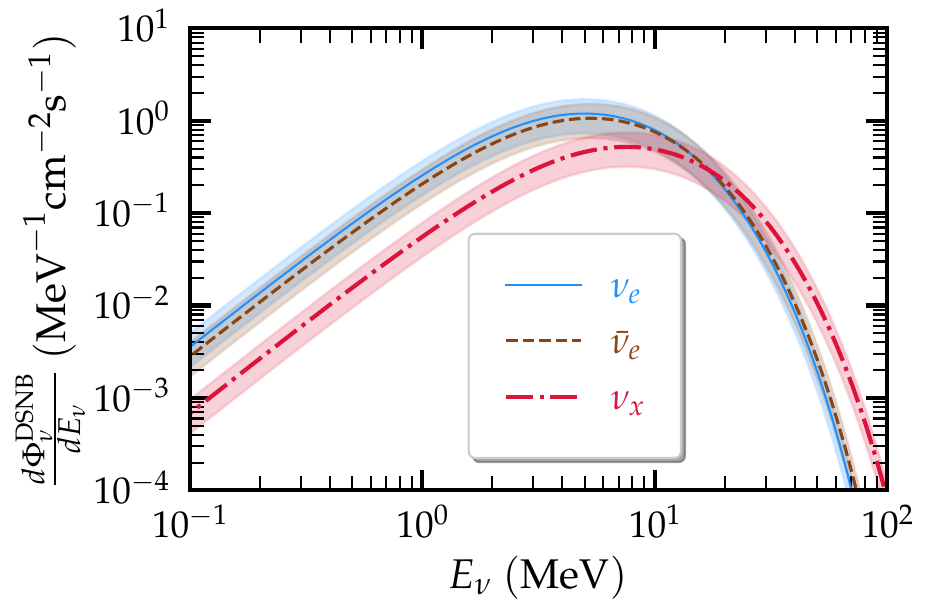}
 \caption{Predicted DSNB fluxes for various neutrino flavors ($\nu_x$ denotes either $\nu_\mu$ or $\nu_\tau$ or their antiparticles), as a function of the neutrino energy, estimated from Eq.~\eqref{equn:DSNB_diff_Flux}. The bands illustrate a 40\% error in the normalization uncertainty of the DSNB spectra~\cite{Horiuchi:2008jz}.}
 \label{fig:DSNB_Flux}
 \end{figure}

We show in Fig.~\ref{fig:DSNB_Flux} the predicted DSNB fluxes, for the different neutrino flavors, as a function of the neutrino energy. In the following calculations we will assume an uncertainty of 40\% in the normalization of the DSNB spectra, estimated from uncertainties in the cosmic star-formation rate~\cite{Horiuchi:2008jz}.
This uncertainty on the DSNB fluxes is illustrated by the shaded bands in Fig.~\ref{fig:DSNB_Flux}. 

\section{\label{Sec:Boosted_DM}The DSNB-boosted dark matter flux}

In this section, we discuss how the DM particles in the Milky Way halo get boosted to considerably greater velocities due to their scattering with DSNB neutrinos. We remain agnostic of the specific DM model\footnote{For a model-dependent study exploring light mediators but focusing on cosmic-rays boosted DM see e.g.  Ref.~\cite{Bell:2023sdq}.} and for the sake of uniformity in comparing the final results we assume the DM to be made of one particle species $\chi$ that scatters with neutrinos and electrons ($\sigma_{\nu \chi}=\sigma_{\chi e}$) or with neutrinos and nucleons ($ \sigma_{\nu \chi} =\sigma_{\chi n}$) with the same cross section,  as the benchmark for our analysis.
These assumptions can be naturally realized in flavor-dependent gauged $U(1)$ extensions such as $U(1)_{B_i - 3L_i}$, $i$ being generation index or $U(1)_{B-3L_{i}}$ models~\cite{Ma:1997nq, Bonilla:2017lsq,delaVega:2021wpx} (for an overview of such a model, see Appendix~\ref{appendixE}). Furthermore, scenarios deviating from this assumption can be easily accounted for by using the product $\sqrt{\sigma_{\nu \chi} \sigma_{\chi e}}$ or $\sqrt{\sigma_{\nu \chi} \sigma_{\chi n}}$ as applicable, see Sec.~\ref{sec:results} for further discussion. Before entering into details, it is noteworthy to stress that the initial DM galactic escape velocity is irrelevant~\cite{Bringmann:2018cvk},  as the scattering between $\chi$ and  DSNB neutrinos accelerates the DM to significantly higher velocities.

The DSNB-boosted DM differential flux, induced by its scattering with the DSNB given in Eq.~\eqref{equn:DSNB_diff_Flux}, can be estimated as~\cite{Das:2021lcr}

\begin{equation}
\label{equn:Boosted_DM_Flux}
\frac{d\Phi_\chi}{dT_\chi} = D_\mathrm{halo}\int_{E_\nu^\mathrm{min}}^{E_\nu^\mathrm{max}}dE_\nu \frac{1}{m_\chi}\frac{d \sigma_{\nu \chi}}{d T_\chi} \frac{d\Phi_\nu^\mathrm{DSNB}}{dE_\nu}\,,
\end{equation}
where $T_\chi$ is the energy transferred to $\chi$ and $\frac{d\Phi_\nu^\mathrm{DSNB}}{dE_\nu}$ is the sum over all neutrino flavors of the DSNB flux given in Eq.~\eqref{equn:DSNB_diff_Flux}. The neutrino-DM scattering cross section  can be cast in the form
\begin{equation}
\frac{d \sigma_{\nu \chi}}{d T_\chi}= \frac{\sigma_{\nu\chi}}{T_\chi^\mathrm{max}(E_\nu)} \Theta\left[T_\chi^\mathrm{max}(E_\nu)-T_\chi\right] \, ,
\end{equation}
where $m_\chi$ denotes the DM mass, while $\sigma_{\nu \chi}$ controls the strength of the interaction. The maximum transferred energy to which the DM can be boosted for a given neutrino energy $E_\nu$, is dictated by the kinematics of the process and is incorporated in the Heaviside step function~\footnote{Throughout our study, we have taken neutrinos as massless since corrections due to non-zero neutrino masses are very small.}: $T_\chi^\mathrm{max}(E_\nu)=E_\nu^2\Big/\Big(E_\nu+\frac{m_\chi}{2}\Big)$. The maximum neutrino energy in our numerical calculations is taken to be $E_\nu^\mathrm{max}=100$~MeV, while the lower integration limit in Eq.~(\ref{equn:Boosted_DM_Flux}) can be obtained by inverting the expression for $T_\chi^\mathrm{max}$ which gives the minimum neutrino energy required to boost the DM, i.e. $E_\nu^\mathrm{min}=\Big[T_\chi+\sqrt{T_\chi^2+2m_\chi T_\chi}\mathrm{~}\Big]\Big/2$.

The $D-$factor ($D_\mathrm{halo})$ in Eq.~(\ref{equn:Boosted_DM_Flux}) encodes the DM density distribution within our galactic halo, and it is expressed as the integral of the density profile along the line of sight (l.o.s.) $\ell$ and over the solid angle $\Omega$: 
\begin{equation}
\label{eq:Dfactor}
D_\mathrm{halo}= \int_{\Delta \Omega} \frac{d\Omega}{4 \pi} \int_0^{\ell_\mathrm{max}} \rho_\mathrm{MW}[r(\ell,\psi)]d\ell .
\end{equation}
Here, we assume a Navarro-Frenk-White (NFW) profile\footnote{The simulated events are found to be largely independent of the DM density profile. We have checked that using a cored isothermal DM density profile~\cite{Ng:2013xha} $D_\mathrm{halo}$ changes by less than $1\%$.}, 
defined as~\cite{Navarro:1996gj}
\begin{equation}
\label{equn:NFW}
\rho_\mathrm{MW}(r)=\rho_\odot\left[\frac{r}{r_\odot}\right]^{-1}\left[\frac{1+\frac{r_\odot}{r_s}}{1+\frac{r}{r_s}}\right]^2\,,
\end{equation}
where the scale radius is $r_s=20~\mathrm{kpc}$ and the local DM density is $\rho_\odot=0.4~\mathrm{GeV~cm}^{-3}$.
The galactocentric distance reads 
\begin{equation}
 r(l,\psi)=\sqrt{r_\odot^2-2lr_\odot\cos{\psi}+l^2} \, ,   
\end{equation}
with $r_\odot=8.5~\mathrm{ kpc}$ being the distance between the Earth and the galactic centre
and $\psi$ the angle of view defining the l.o.s.. 
The upper limit of the l.o.s. integral is given by $\ell_\mathrm{max}=\sqrt{R^2-r_\odot^2\sin^2{\psi}}+r_\odot\cos{\psi}$, with the galactic halo virial radius taken to be $R=200~\mathrm{ kpc}$. 
Given these values, we hence obtain $D_\mathrm{halo}=2.22 \times 10^{25}\mathrm{~MeV~cm}^{-2}$ over the whole galactic halo.

\subsection{Attenuation effects}
\label{sec:attenuation}

In this subsection, we will focus our attention on the modifications expected to occur in the energy profile of the DSNB-boosted DM flux during its propagation through the atmosphere and the Earth~\cite{Kavanagh:2016pyr,Mack:2007xj,Emken:2018run,Starkman:1990nj,Bringmann:2018cvk,Xia:2021vbz}.
For sufficiently large interaction cross sections, $d \sigma_{\chi i}/d T_i$, the DM particles may lose a significant amount of energy due to their scattering on nuclei ($i=\mathcal{N}$) or electrons ($i=e$), resulting into a sizeable attenuation of the DM flux before reaching the detector. 
This effect can be accounted for via the energy loss equation~\cite{Bringmann:2018cvk,Xia:2021vbz}
\begin{equation}
\label{equn:Attenuation}
\begin{split}
\frac{dT_\chi^z}{dz}=&-n_i\int_0^{T_i^\mathrm{max}(T_\chi^z)} \frac{d \sigma_{\chi i}}{d T_i} T_i \, dT_i\,,
\end{split}
\end{equation}
where $T_i$ denotes the energy lost by the boosted DM particle in a collision and  $n_i$ is the number density of nucleus species or  electrons. Here, $z$ denotes the distance travelled from the location of the scattering point (inside the atmosphere or the Earth) to the detector.
In the most general case, Eq.~(\ref{equn:Attenuation}) relates the initial energy at the top of the atmosphere ($z=0$), $T_\chi^0$, with the average kinetic energy, $T_\chi^z$, after travelling a distance $z$ before reaching the underground detector. In our analysis, we  neglect the impact of atmospheric attenuation as it is expected to be negligible compared to  Earth's attenuation~\cite{Xia:2021vbz}, for the cross sections under consideration. Hence, we take $z=0$ at the Earth's surface. Then, the 
distance $z$ can be expressed as~(see Appendix~\ref{appendix1} for more details)
\begin{equation}
\label{equn:los}
z=-(R_E-h_d)\cos{\theta_z}+\sqrt{R_E^2-(R_E-h_d)^2\sin^2{\theta_z}}\,,
\end{equation}
where $R_E$ stands for the radius of the Earth, $\theta_z$ refers to the detector's zenith angle and $h_d$ indicates the depth of the detector's location from the Earth's surface, at the point where the zenith angle is zero. 
 Moreover, for the sake of simplicity, we have adopted a mean average electron density $n_e$ of Earth’s most abundant elements between the surface and depth $z$,
 $n_e=8\times 10^{23} \mathrm{~cm}^{-3}$~\cite{Ema:2018bih}. 
In the case of attenuation due to $\chi$ scattering on nuclei, we have determined the nuclear number density at depth $z$
through a weighted average of the
most abundant elements  found in the Earth’s crust, mantle, and core, yielding $n_\mathcal{N} =  3.44 \times 10^{22} \mathrm{~cm}^{-3}$  and  $A \approx33.3$ (for details see Appendix~\ref{appendixB} and Refs.~\cite{morgan, MCDONOUGH2003547, Dziewonski:1981xy, Kavanagh:2016pyr}). 

The differential cross section for DM-electron or DM-nucleus scattering takes the form
\begin{equation}
\label{eq:DM-nucleus-electron}
    \frac{d \sigma_{\chi i}}{d T_i} = \frac{\sigma_{\chi i}}{T_i^\text{max} (T_\chi)} \, , \quad i= \mathcal{N}~\text{or}~e\, ,
\end{equation}
where the maximum recoil energy that can be lost by $\chi$ during the attenuation process, is obtained from the kinematics of the process and reads
\begin{equation}
\label{equn:Max_Recoil_Energy_Detector}
T_i^\mathrm{max}(T_\chi)=\frac{2m_iT_\chi(T_\chi+2m_\chi)}{(m_\chi+m_i)^2+2m_iT_\chi}\, ,
\end{equation}
with $m_i$ indicating the nuclear ($i = \mathcal{N}$) or electron ($i = e$) mass.
The solution of Eq.~(\ref{equn:Attenuation}) gives the DM energy as a function of the distance and the initial DM energy, i.e. $T_\chi^z \equiv T_
\chi^z (T_\chi^0, z)$ with $z$ depending on the zenith angle and the detector depth as indicated in Eq.~(\ref{equn:los}). The resulting attenuated DM flux reaching the detector after averaging over angles~\footnote{Here, we only consider the angle-averaged DM flux as most of the current
DM direct detection experiments do not have directionality capabilities.}, $d \Phi_\chi^z/ dT_\chi^z$, is given by the expression 
\begin{equation}
\label{equn:BDM_Flux_General}
\frac{d\Phi_\chi}{dT_\chi^z}=\int d\Omega \left.\frac{d^2\Phi_\chi}{dT_\chi d\Omega}\right|_{T_\chi^0} \frac{dT_\chi^0}{dT_\chi^z}\, ,
\end{equation}
where $\Omega$ is the solid angle. 

\subsubsection{Scattering with electrons}
\label{sect.:electron-attenuation}

For the case of DM-electron scattering we have $\sigma_{\chi i}= \sigma_{\chi e}$ in Eq.~(\ref{eq:DM-nucleus-electron}). In this case Eq.~\eqref{equn:Attenuation} can be solved analytically. The solution for $T_\chi^z$ at a given depth $z$ can be expressed in terms of the DM energy at the surface, $T_\chi^0$, as
\begin{equation}
    \label{equn:Txz_DM_e}
    T_\chi^z\approx\frac{T_\chi^0e^{-z/l_E}}{1+\frac{T_\chi^0}{2m_\chi}\left(1-e^{-z/l_E}\right)}\, ,
\end{equation}
where $l_E$ represents the mean free path for energy loss, given by $l_E^{-1}=n_e\sigma_{\chi e}\frac{2m_em_\chi}{(m_e+m_\chi)^2}$. By inverting Eq.~\eqref{equn:Txz_DM_e} we obtain the expression for $T_\chi^0$ as a function of $T_\chi^z$ and $z$, which reads
\begin{equation}
    T_\chi^0\approx \frac{2m_\chi T_\chi^z e^{z/l_E}}{2m_\chi+T_\chi^z\left(1-e^{z/l_E}\right)}\, .
\end{equation}
As a consequence, the attenuated DM flux given in Eq.~\eqref{equn:BDM_Flux_General} that eventually reaches the detector can be simplified as follows
\begin{equation}
\label{eq:attenuated-flux-electron}
    \frac{d\Phi_\chi}{dT_\chi^z}\approx \int d\Omega \left.\frac{d^2\Phi_\chi}{dT_\chi d\Omega}\right|_{T_\chi^0} \frac{4m_\chi^2e^{z/l_E}}{\left[2m_\chi+T_\chi^z\left(1-e^{z/l_E}\right)\right]^2}\, .
\end{equation}

Before closing this discussion let us stress that, as discussed before, our analysis is done for the benchmark $\sigma_{\nu\chi}= \sigma_{\chi e}$. Note that the bounds that we will eventually obtain in Sec.~\ref{sec:results} will mainly depend on the product $\sqrt{\sigma_{\nu\chi} \sigma_{\chi e}}$, which simplifies to $\sigma_{\chi e}$ under the assumption $\sigma_{\nu\chi}= \sigma_{\chi e}$, plus corrections due to the dependence of $l_E$ (see Eq.~\ref{equn:Txz_DM_e}) on $\sigma_{\chi e}$. We will discuss this in more detail in Sec.~\ref{sec:results}. 

\subsubsection{Scattering with nuclei: Effect of the finite nuclear size}
\label{sect.:nucleus-attenuation}

 For the case of DM-nucleus scattering we take $\sigma_{\chi i} = \sigma_{\chi \mathcal{N}}^\text{SI}$ in Eq.~(\ref{eq:DM-nucleus-electron}), where the spin-independent (SI) DM-nucleus elastic scattering cross section is expressed as~\footnote{For simplicity we consider a spin-conserving scenario where DM-proton and DM-neutron effective couplings are equal ($f_p/f_n\approx1$).}~\cite{Lewin:1995rx, Freese:2012xd}
\begin{equation}
\label{equn:DM-N_SI_xSec}
\sigma_{\chi \mathcal{N}}^\text{SI}(\textgoth{q}^2)=\frac{\mu_\mathcal{N}^2}{\mu_n^2}A^2\sigma_{\chi n} F^2(\textgoth{q}^2)\,,
\end{equation}
where $A$ denotes the atomic mass number of the target nuclei, $\textgoth{q} = \sqrt{2 m_\mathcal{N} T_\mathcal{N}}$ stands for the three-momentum transfer, $\sigma_{\chi n}$ is the DM-nucleon SI cross section and $\mu_\mathcal{N}$ ($\mu_n$) represents the DM-nucleus (DM-nucleon) reduced mass.  We again take the benchmark $\sigma_{\nu \chi} = \sigma_{\chi n}$ for our analysis, but, as mentioned above, our bounds are mainly dependent on $\sqrt{\sigma_{\nu \chi} \sigma_{\chi n}}$ (see Sec.~\ref{sec:results}). The SI cross section $\sigma_{\chi \mathcal{N}}^\text{SI}$ is a momentum-dependent quantity due to the presence of the  nuclear form factor, $F(\textgoth{q}^2)$, which accounts for the finite nuclear size and has been parametrized by a Helm-type~\footnote{The results are independent from the choice of the nuclear form factor; other well-known parametrizations such as the Klein-Nystrand and Symmetrized Fermi will lead to the same conclusions.} effective form factor~\cite{Helm:1956zz}. Notice that the energy dependence of the SI cross section prevents us from obtaining an analytical solution for Eq.~(\ref{equn:Attenuation}), unlike the case of DM-electron scattering discussed above, hence we need to solve Eq.~(\ref{equn:Attenuation}) numerically.

\begin{figure}[t]
 \includegraphics[width=0.53 \textwidth]{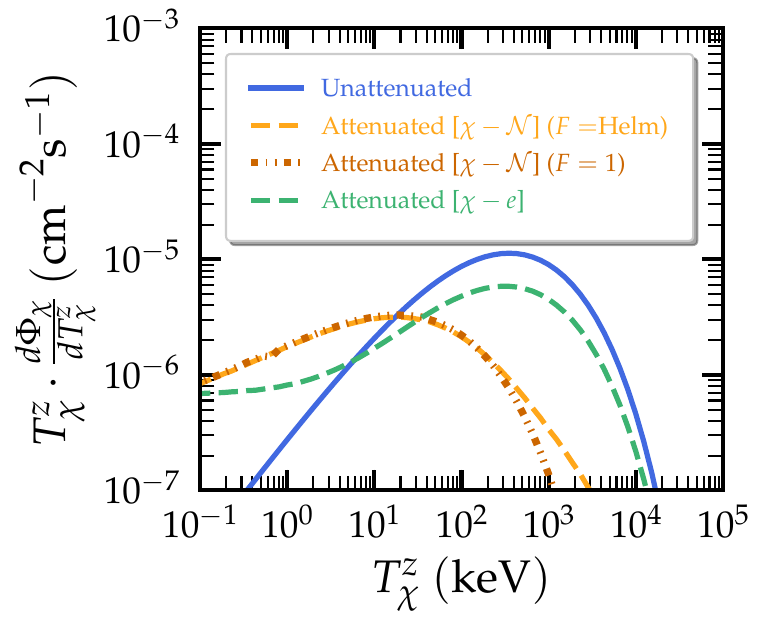}
 \caption{The angle-averaged DNSB-boosted DM flux distribution as a function of the DM energy for $m_{\chi} = 300 \, \mathrm{MeV}$,  $\sigma_{\nu \chi} = 10^{-29} \, \mathrm{cm}^2$ and a  detector depth of $h_d = 1.4$~km for the benchmarks $\sigma_{\nu \chi} = \sigma_{\chi e}$ and $ \sigma_{\nu \chi} = \sigma_{\chi n}$. The unattenuated flux is shown by the solid blue line. The attenuated DM flux for the case of DM scattering with electrons (nuclei including Helm-type form factor)  inside the Earth is displayed with a green (orange) dashed line. The effect of the nuclear form factor is illustrated by comparing the results assuming a Helm-type form factor (dashed line) and $F=1$ (dash-dotted line). See main text for more details.}
 \label{fig:BDM_Flux}
 \end{figure}

The fact that the DM travels through Earth to reach the detectors, leads to attenuation of the flux. In Fig.~\ref{fig:BDM_Flux} we present the angle-averaged  DSNB-boosted DM flux for unattenuated and attenuated cases. The results are plotted for the benchmark parameters $m_{\chi} = 300 \, \mathrm{MeV}$ and $\sigma_{\nu \chi} = 10^{-29} \, \mathrm{cm}^2$,  assuming a depth of $h_d=1.4$~km which corresponds to the underground location of XENONnT. Note that the results remain essentially unchanged for the case of LZ ($h_d=1.47$~km).
The solid blue line in Fig.~\ref{fig:BDM_Flux} corresponds to the unattenuated flux. The dashed lines show the attenuated fluxes corresponding to attenuation due to DM-nucleus (orange, dashed) and DM-electron (green, dashed) scattering as a function of the DM energy. For the case of $\chi-\mathcal{N}$ scattering, the effect of the finite nuclear size is illustrated by comparing the resulting fluxes for two cases: i) by incorporating the Helm form factor in the calculation (orange, dashed) and ii) by assuming $F=1$  i.e. completely ignoring nuclear physics (orange, dotted).

As can be seen, Earth's attenuation effects shift the peak of the DSNB-boosted DM flux towards lower energies and reduce it up to a factor 2 (3.5) for $\chi-e$ ($\chi-\mathcal{N}$) scattering. Furthermore, the high-energy endpoint of the differential DSNB-boosted DM flux spectra exhibits a faster decline when finite nuclear size is neglected, as opposed to the case where the finite nuclear size effects are taken into account. The Earth's attenuation and nuclear size effects play a crucial role in the results presented in the remainder of the work (see also Appendix~\ref{appendix2}). 

Before closing this discussion let us stress that in the present analysis, the calculated attenuated flux is not taking into account additional effects related to the direction of DNSB-boosted DM particles
after each scattering process nor the possibility of multiple scatterings. These effects only become relevant when the DM mass and energy are significantly lower than
the mass of the nucleus~\cite{Xia:2021vbz, Kavanagh:2016pyr} and are important for probing diurnal effects~\cite{Ge:2020yuf}.

\section{\label{Sec:DM_Scattering}Dark matter signal at underground detectors}

After reaching the underground detector, the DSNB-boosted DM can scatter off both the electrons and nuclei of the target material, thus inducing both electronic and nuclear recoils. The differential event rate with respect to the recoil energy $T_i$ can be written as~\cite{Das:2021lcr}
\begin{equation}
\label{equn:Diff_Events_Rate}
\frac{dR}{dT_i}=t_\mathrm{run}N_\mathrm{target}^i \mathcal{A}\int dT_\chi^z\frac{d\Phi_\chi}{dT_\chi^z}\frac{\sigma_{\chi i}}{T_i^\mathrm{max}(T_\chi^z)}\Theta[T_i^\mathrm{max}(T_\chi^z)-T_i]\,,
\end{equation}
where $t_\mathrm{run}$ and $N_\mathrm{target}$ denote the exposure time and number of targets of the detector, respectively, while $\mathcal{A}$ represents the detection efficiency provided by each experiment. At this point we should clarify that the detection efficiency is provided either in terms of true $\mathcal{A}(T_i)$ or reconstructed $\mathcal{A}(T_i^\mathrm{reco})$ recoil energy, hence its explicit dependence has been dropped in Eq.~\eqref{equn:Diff_Events_Rate} to avoid confusion. Regarding DM-electron scattering, our calculations incorporate the detection efficiency provided by LZ~\cite{LZ:2023poo} in terms of true recoil energy $\mathcal{A}(T_e)$, while for the case of XENONnT we consider the efficiency provided in Ref.~\cite{XENON:2022ltv} in terms of reconstructed recoil energy $\mathcal{A}(T_e^\mathrm{reco})$. Regarding DM-nucleus scattering, we account for the detection efficiency $\mathcal{A}(T_\mathcal{N})$ provided in terms of true nuclear recoil energy as reported by both LZ~\cite{LUX-ZEPLIN:2022xrq} and XENONnT~\cite{XENONCollaboration:2023orw}.
In what follows, we will use the recent data released by XENONnT and LZ collaborations to put constraints on the DM mass and DM-electron/nucleon cross sections. 

We first focus on DM-electron scattering, i.e. we calculate the differential event spectrum $dR/dT_e$ given in Eq.~(\ref{equn:Diff_Events_Rate}) for $i=e$. In this case, $\chi$ particles scatter off electrons in the underground detector with a cross section $\sigma_{\chi e}$. The angle-averaged DSNB-boosted DM flux, accounting for the attenuation effects (see Sec.~\ref{sect.:electron-attenuation}) is given in Eq.~(\ref{eq:attenuated-flux-electron}). Since very low energy scatterings occur, our calculations take into consideration atomic binding effects which lead to a slight cross section suppression at very low recoil energies. To this purpose, the number of target electrons in Eq.~(\ref{equn:Diff_Events_Rate}) is expressed as $N^e_\text{target}= \frac{m_\text{det} N_A}{M_r}\times Z_\text{eff}(T_e)$ where $m_\text{det}$, $M_r$ and $N_A$ represent the detector mass, the molar mass of the target material and the Avogadro's number, respectively. The recoil energy-dependent quantity $Z_\text{eff}(T_e)$ denotes the effective charge of the atomic nucleus that is seen by DM for a given energy deposition $T_e$. The latter can be approximated by a series of step functions that depend on the single particle binding energy of the $i$th electron, following the Hartree-Fock calculations of Ref.~\cite{Chen:2016eab}.

Turning our attention to the case of DM scattering with nuclei, we calculate the corresponding differential event spectrum $dR/dT_\mathcal{N}$ that follows from Eq.~(\ref{equn:Diff_Events_Rate}) for $i=\mathcal{N}$. For the sake of clarity let us note that in this case, although our calculated event rates refer to DM-nucleus scattering, our results will be always expressed in terms of the fundamental DM-nucleon cross section $\sigma_{\chi n}$, instead of the SI DM-nucleus cross section $\sigma^\text{SI}_{\chi \mathcal{N}}$ [see e.g. Eq.~\eqref{equn:DM-N_SI_xSec}]. In this case the angle-averaged attenuated boosted DM flux has been computed numerically as discussed in Sec.~\ref{sect.:nucleus-attenuation}.
Since both the LZ and XENONnT collaborations  have reported their measured data in terms of electron-recoil spectra,  we convert our calculated nuclear recoil spectrum $dR/dT_\mathcal{N}$ into an ``electron-equivalent'' recoil spectrum, according to the expression
\begin{equation}
    \frac{dR}{dT_e} = \frac{dR}{dT_\mathcal{N}}  \frac{1}{\mathcal{Q}_f(T_\mathcal{N}) + T_\mathcal{N} \frac{d \mathcal{Q}_f}{d T_\mathcal{N}}} \, ,
\end{equation}
where the quenching factor, $\mathcal{Q}_f(T_\mathcal{N}) = T_e/T_\mathcal{N}$, quantifies the energy loss to heat in the aftermath of a DM-nucleus scattering event. In the present analysis we adopt the standard Lindhard quenching factor~\cite{Lindhard:1963}. Notice also that for DM-nucleus scattering the effective charge $Z_\text{eff}$ is irrelevant and hence we take $Z_\text{eff}=1$.

 In the final step we intend to simulate the expected boosted DM signal at the LZ and XENONnT detectors with high reliability. To this purpose we evaluate the reconstructed spectra by assuming a Gaussian  smearing function $\mathcal{G}(T_e^\text{reco}, T_e)$ as
\begin{equation}
\label{eq:Reco_Spectrum}
    \frac{d R}{d T_e^\text{reco}} = \int_{0}^{T_e^\text{max}} \frac{dR}{d T_e}(T_e) \mathcal{G}(T_e^\text{reco}, T_e) \, d T_e \, . 
\end{equation}
By following the above method we have verified that our predictions are in excellent agreement with those reported by XENONnT and LZ for both electron and nuclear recoils. First, based on previous work~\cite{A:2022acy}, we have calculated the elastic neutrino-electron scattering spectra and found that a total of $\sim$30~events are expected for LZ and 76~events (300~events in the full region [1,140] $\mathrm{keV_{ee}}$) for XENONnT, in agreement with Refs.~\cite{LUX-ZEPLIN:2022xrq, XENON:2022ltv}. Second, regarding nuclear recoils we have calculated the corresponding coherent elastic neutrino-nucleus scattering (CE$\nu$NS) expected events induced by $^8$B neutrinos and found 0.16~events  for LZ and 0.24~events for XENONnT, in agreement with Refs.~\cite{LUX-ZEPLIN:2022xrq, XENONCollaboration:2023orw}, respectively.

\begin{figure}[t]
 \includegraphics[width=0.48 \textwidth]{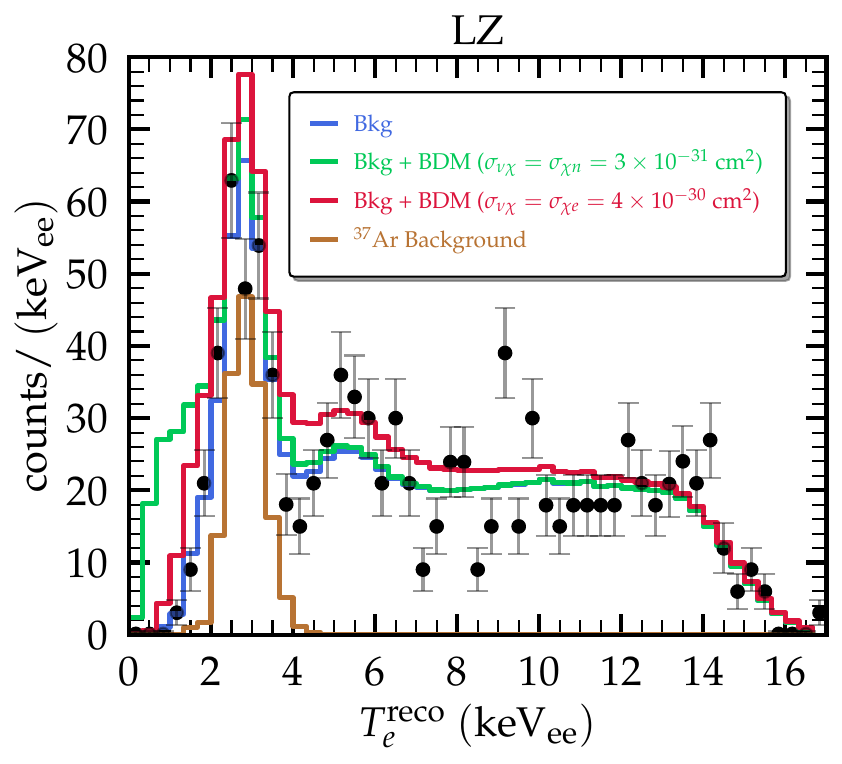}
 \includegraphics[width=0.49 \textwidth]{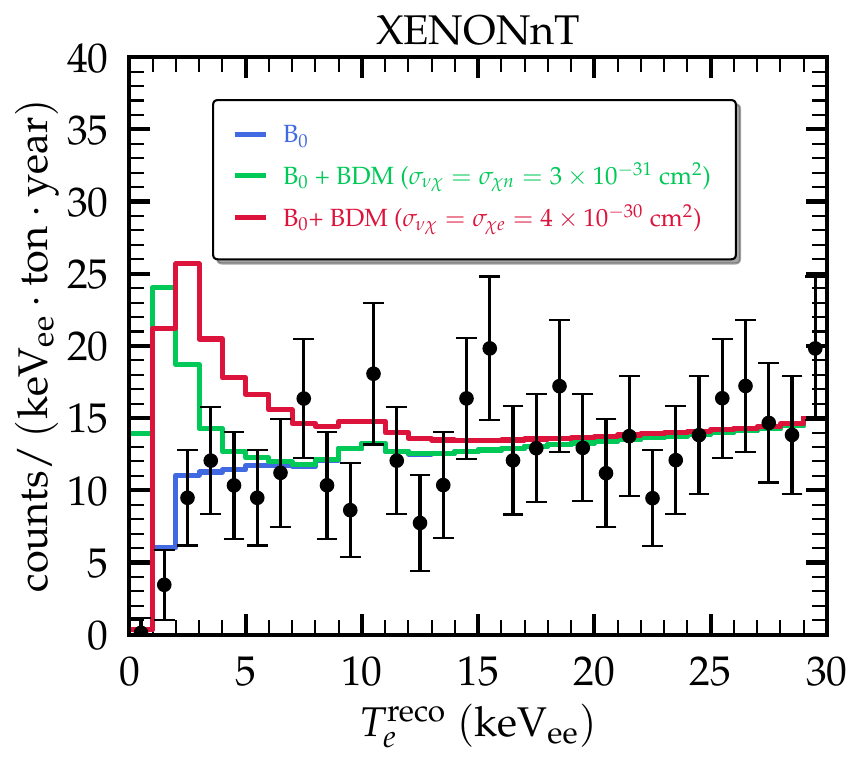}
 \caption{Simulated signal (colored histograms) and experimental data (black points with error bars) as a function of the electron-equivalent ionization energy, for the LZ (left) and XENONnT (right) experiments. The DSNB-boosted DM events have been computed assuming $m_\chi=300\mathrm{~MeV}$ and different values of scattering cross sections, see the legend. The blue and light brown histograms refer to the background events as provided by the collaborations.}
 \label{fig:events}
 \end{figure}

Following these prescriptions, we show in Fig.~\ref{fig:events} the simulated recoil spectra as a function of the electron-equivalent ionization energy, expected at both the LZ and XENONnT detectors. The black points depict the experimental data together with the error bars, when available~\footnote{The LZ collaboration does not provide errors in their reported experimental data, thus we assume  standard errors.}, as provided by the collaborations. The blue histograms indicate the background events, also given by the collaborations. The red and green histograms represent the sum of the background and of the simulated number of events, assuming $m_\chi=300\mathrm{~MeV}$ and two different values of $\chi$-nucleon or $\chi$-electron scattering cross sections, as indicated in the legend. In the case of the LZ detector, the light brown histogram further represents the $^{37}$Ar background, which originates from cosmogenic activation of the xenon prior to underground deployment, producing short-lived $^{37}$Ar that decayed during the first run~\cite{LUX-ZEPLIN:2022xrq}.

%
%
\subsection{LZ analysis}

For the analysis of LZ data, we have performed a spectral analysis using the following Poissonian $\chi^2$ function~\cite{Almeida:1999ie}
\begin{equation}
\begin{aligned}
\chi^2(\overrightarrow{\mathcal{S}}; \alpha,\beta,\delta)=&2\sum_{i=1}^{51} \Biggl[  R_\mathrm{pred}^i(\overrightarrow{\mathcal{S}}; \alpha,\beta,\delta) - R_\mathrm{exp}^i+  R_\mathrm{exp}^i\ln \left(\frac{R_\mathrm{exp}^i}{ R_\mathrm{pred}^i(\overrightarrow{\mathcal{S}}; \alpha,\beta,\delta)}\right) \Biggr]+\\&\left(\frac{\alpha}{\sigma_\alpha}\right)^2+\left(\frac{\beta}{\sigma_\beta}\right)^2+\left(\frac{\delta}{\sigma_\delta}\right)^2 \, ,
\end{aligned}
\label{equn:poisonian_chi_square}
\end{equation}
where $R_\mathrm{exp}^i$ denotes the experimental differential events in the $i$th recoil energy bin, as reported in Ref.~\cite{LUX-ZEPLIN:2022xrq}, while the predicted differential events contain the DSNB-boosted DM signal, as well as all background components: $R_\mathrm{pred}^i(\overrightarrow{\mathcal{S}}; \alpha,\beta,\delta)=(1+\alpha)R_\mathrm{bkg}^i+(1+\beta)R_\text{DSNB-BDM}^i(\overrightarrow{\mathcal{S}})+(1+\delta)R_{^{37}\text{\text{Ar}}}^i$. It is worth noting that, in accordance with Ref.~\cite{AtzoriCorona:2022jeb}, the $R_\mathrm{bkg}$ spectrum is calculated by eliminating the $^{37}\mathrm{Ar}$ contributions from the total background provided in Ref.~\cite{LUX-ZEPLIN:2022xrq}. The nuisance parameters $\{\alpha,\beta,\delta\}$ are introduced to incorporate the uncertainty on  background, neutrino flux distribution\footnote{The uncertainty on the DSNB flux primarily arises from the uncertainty in the star-formation rate, mentioned in Eq.~\eqref{equn:SFR}.} and $^{37}\text{Ar}$ components with $\sigma_\alpha=13\%,\mathrm{~}\sigma_\beta=40\%\text{~\cite{Horiuchi:2008jz} and }\sigma_\delta=100\%$. For each new physics parameter belonging to $\overrightarrow{\mathcal{S}}$ (i.e. $m_\chi$ or $\sigma_{\chi i}$), we have marginalized the $\chi^2$ function over all nuisance parameters.

\subsection{XENONnT analysis}

The following Gaussian $\chi^2$ function is used for the analysis of XENONnT data~\cite{Almeida:1999ie}
\begin{equation}
\label{equn:gaussian_chi_square}
\chi^2(\overrightarrow{\mathcal{S}}; \beta)= \sum_{i=1}^{30} \left( \frac{R_\text{pred}^i(\overrightarrow{\mathcal{S}};\beta)-R_\text{exp}^i}{\sigma^i} \right)^2 + \left( \frac{\beta}{\sigma_\beta}\right)^2 \, .
\end{equation}
Here,  $R_\text{pred}^i(\overrightarrow{\mathcal{S}}; \beta)=\left(1+\beta \right) R^{i}_\text{DSNB-BDM}(\overrightarrow{\mathcal{S}}) + B_0^i$, with $B_0$ denoting the simulated background mentioned in Ref.~\cite{XENON:2022ltv}. The rest of the details are similar to the LZ analysis.

\section{Results}
\label{sec:results} 

\begin{figure}[ht!]
 \includegraphics[width=0.49 \textwidth]{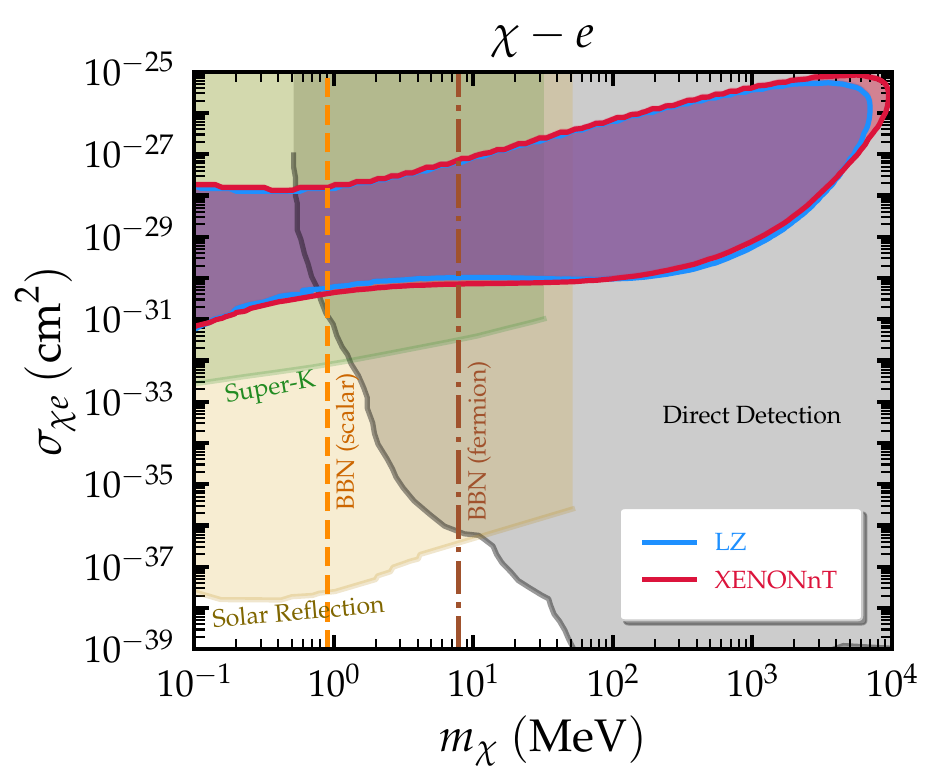}
 \includegraphics[width=0.49 \textwidth]{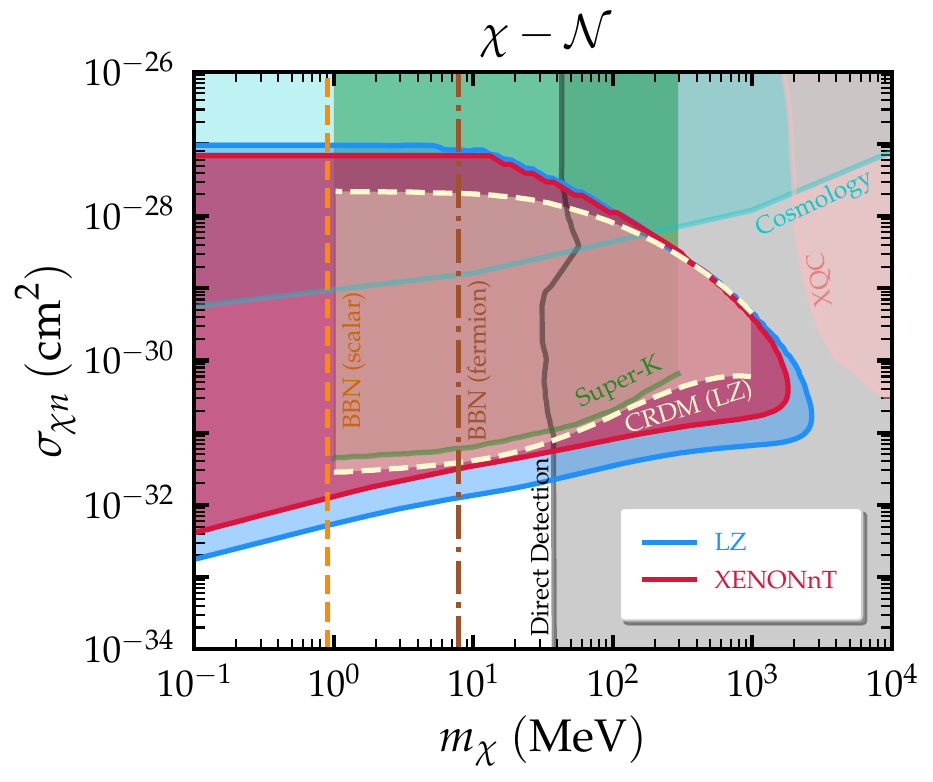}
 \caption{90\% C.L. exclusion regions in the DSNB-boosted DM parameter space, obtained for scattering off electrons, for the benchmark $\sigma_{\nu\chi} = \sigma_{\chi e}$ (left panel) and nuclei, for the benchmark $\sigma_{\nu\chi} = \sigma_{\chi n}$ (right). We show results obtained with LZ (blue) and XENONnT (red) data. For comparison purposes, existing limits from other studies are also shown (see main text for details).}
 \label{fig:param_space}
 \end{figure}

 We present in Fig.~\ref{fig:param_space} the 90\% C.L. exclusion regions on the DSNB-boosted DM, in the  planes $(m_\chi, \sigma_{\chi e})$ and $(m_\chi, \sigma_{\chi n})$. We consider the case of DM scattering off electrons (left panel) and nuclei (right panel), and show both constraints obtained using LZ (blue) and XENONnT (red) experimental data.

For comparison purposes, we also show limits from other dedicated experiments and studies. In the $\chi$-electron scattering channel (left panel), we consider results from DD experiments (gray contour), mainly SENSEI~\cite{SENSEI:2020dpa}, DAMIC~\cite{DAMIC:2019dcn}, EDELWEISS~\cite{EDELWEISS:2020fxc}, SuperCDMS~\cite{SuperCDMS:2018mne}, DarkSide-50~\cite{DarkSide:2018ppu}, XENON1T~\cite{XENON:2019gfn} and PandaX-II~\cite{PandaX-II:2021nsg}. Additionally, we include constraints from solar reflection~\cite{An:2017ojc, An:2021qdl} (dark yellow) and from Super-Kamiokande~\cite{Cappiello:2019qsw} (green).
Similarly, for the $\chi$-nucleus scattering channel (right panel), we take into account results from various experiments and studies. These include again DD experiments (black region)~\cite{SuperCDMS:2013eoh, DAMIC:2016lrs, NEWS-G:2017pxg, SuperCDMS:2018mne, CRESST:2019jnq, CDEX:2019hzn, XENON:2019zpr, EDELWEISS:2019vjv, EDELWEISS:2022ktt, DarkSide:2022dhx}, together with constraints from the Super-Kamiokande experiment~\cite{Ema:2020ulo,Super-Kamiokande:2022ncz} (green) and the XQC rocket experiment~\cite{Mahdawi:2018euy} (light pink). We further depict cosmological bounds in cyan (note that they are obtained as a 95$\%$ C.L. exclusion region). Among them, the strongest bounds come from the Lyman-$\alpha$ forest, but also include constraints from cosmic-microwave-background anisotropy measurements and Milky-Way satellite galaxies~\cite{Gluscevic:2017ywp, Slatyer:2018aqg, Xu:2018efh, Nadler:2019zrb, Buen-Abad:2021mvc, Rogers:2021byl}.  We also represent the Big Bang Nucleosynthesis (BBN) limit on the mass of real scalar and Dirac fermion DM~\cite{Krnjaic:2019dzc}. By considering the limits from these experiments, we establish a comprehensive picture of the constraints on the DSNB-boosted DM parameter space. 
Additionally, let us mention that DM-neutrino interactions are in principle subject to strong cosmological bounds. In particular, Lyman-$\alpha$ data provide stringent constraints on $\sigma_{\nu \chi}$~\cite{Wilkinson:2014ksa, Akita:2023yga}, although dependent on several assumptions, including massless neutrinos. A more recent analysis, accounting for neutrino masses,  points toward a preference for a non-zero DM-neutrino interaction strength~\cite{Hooper:2021rjc} thus providing further motivations for our work.

Finally, we compare our results with recent bounds on sub-GeV cosmic-ray boosted DM (CRDM), also derived using LZ data~\cite{Maity:2022exk} (light yellow).
While many references in the literature have addressed cosmic-ray boosted DM, Ref.~\cite{Maity:2022exk} allows for a direct comparison of our results given that we both analyze the same LZ data set. As can be seen, our constraints obtained assuming a boost from DSNB neutrinos are ruling out a larger region of the parameter space. This is understood since the local interstellar population of cosmic rays is about one order of magnitude less intense compared to the flux of DSNB neutrinos, the latter peaking at lower ($\sim 10$ MeV) energies, though. Notice also that Ref.~\cite{Maity:2022exk} ignored nuclear-physics corrections, that are rather important for a CRDM-based analysis where a larger momentum transfer is involved (compared to our DSNB-based analysis).  The correct inclusion of such effects would  drastically modify the CRDM region shown in the plot. Although not shown here, nuclear effects in CRDM studies have been considered by incorporating Helm-type nuclear form factors, for example, in the analysis of XENON1T excess data in Refs.~\cite{Xia:2021vbz, Arguelles:2022fqq}. At this point we should note that given that the initial CRDM flux peaks beyond 100~MeV and extends up to GeV energies, a large momentum transfer is involved  in the process and it cannot be realistically accounted for through the inclusion of nuclear form factors. For an appropriate treatment of nuclear structure at such large momentum transfer see Ref.~\cite{Alvey:2022pad}.

While a significant part of our constraints lie in a region of parameter space already probed by other searches, these results highlight the complementarity and significance of the LZ and XENONnT data in probing the sub-GeV DM parameter space. Also, it is worth mentioning that both these experiments have just started taking data and we are only using their very first data sets obtained with exposure time of only a few months, but still the bounds are already competitive with other bounds. As the statistics of these experiments will increase, their data will play a much more important role in constraining the DSNB-boosted DM parameter space. 
Moreover, and as mentioned in Sec.~\ref{Sec:Boosted_DM}, let us recall that we present the limits on $\sigma_{\chi e}$ and $\sigma_{\chi n}$
under the assumption of $\sigma_{\nu \chi} = \sigma_{\chi e}$ and $\sigma_{\nu \chi} = \sigma_{\chi n}$, respectively. However, note that the lower limit of our closed regions is basically dependent only on $\sqrt{\sigma_{\chi e} \sigma_{\nu \chi}}$ and $\sqrt{\sigma_{\chi n} \sigma_{\nu \chi}}$, respectively, so it can be easily recast into alternative scenarios in which the magnitude of the two cross sections is different. The upper limit of our closed contours though, has a stronger dependence on the attenuation effects and therefore depends on a more complicated combination of $\sigma_{\chi e,n}$ and $\sigma_{\nu \chi}$. 
As it can be noticed, our exclusion regions have a closed shape, due to the inclusion of attenuation effects (see the discussion below). Large scattering cross sections i.e. $\sigma_{\chi e} \gtrsim 2 \times 10^{-28}~\mathrm{cm^2}$ ($\sigma_{\chi n} \gtrsim 8 \times 10^{-28}~\mathrm{cm^2}$) for DM scattering off electrons (nucleons) and $m_\chi = 0.1$~MeV result into a strong attenuation during the propagation of the DM particles through the Earth and are therefore disfavored.

\begin{figure}[h!]
  \includegraphics[width=0.49 \textwidth]{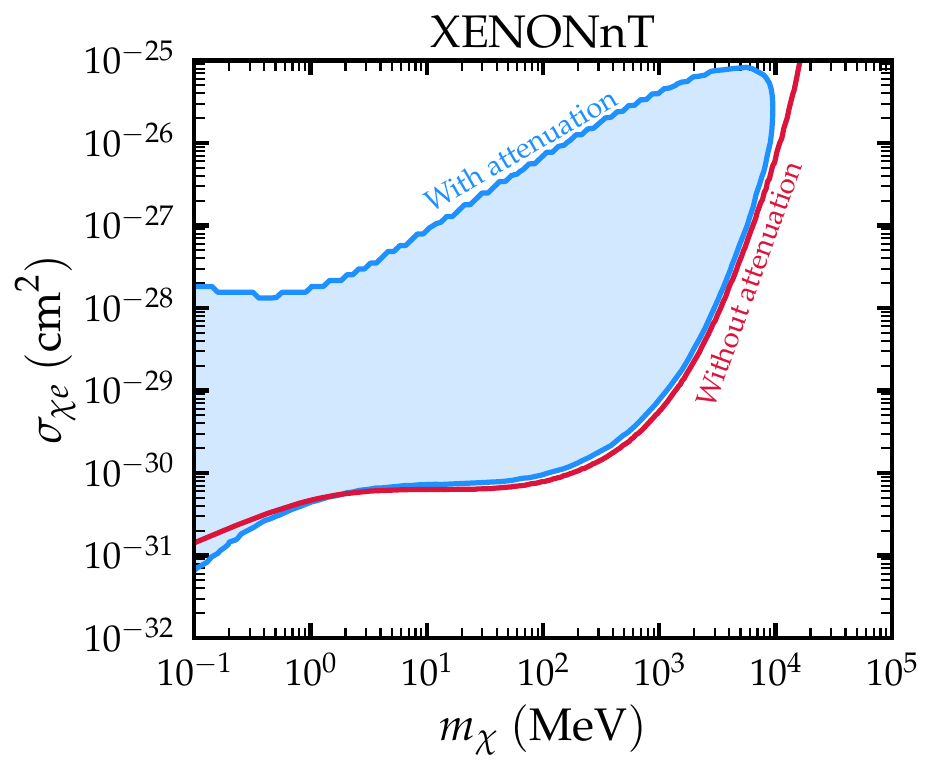}
  \includegraphics[width=0.49 \textwidth]{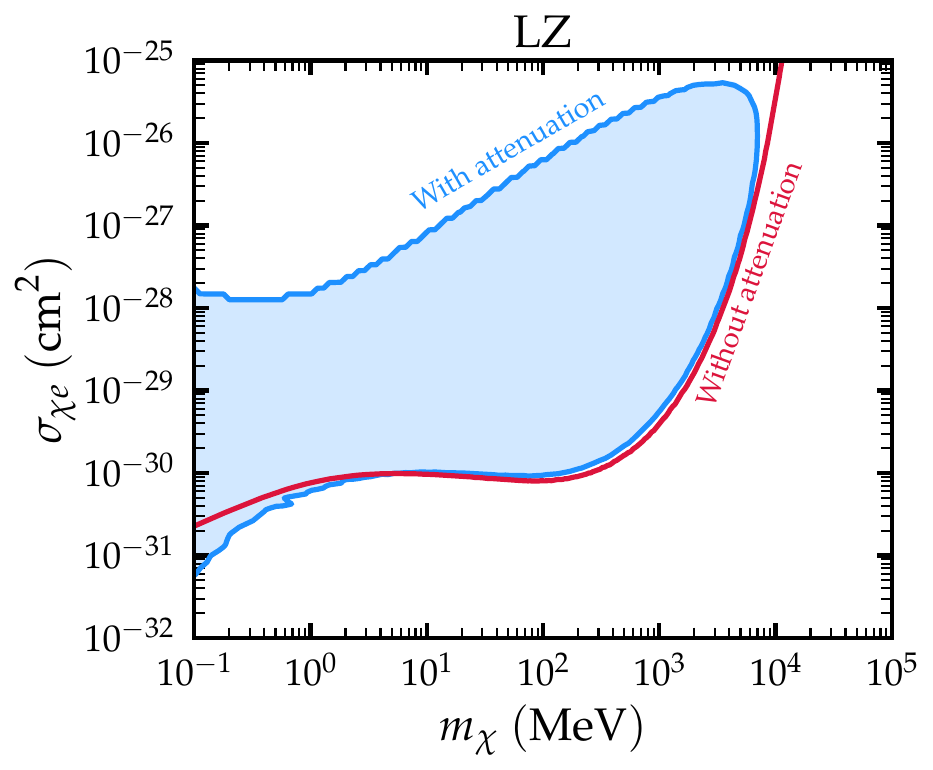}
  \caption{90\% C.L. exclusion limits on $\sigma_{\chi e}$ and $m_\chi$, obtained including (blue closed contour) or neglecting (red line) attenuation effects due to DM scattering off Earth's elements before reaching the detectors. Left panel shows bounds obtained with XENONnT data, while right panel refers to LZ data.} 
 \label{fig:DM_e_Comparison}
 \end{figure}

\begin{figure}[h!]
  \includegraphics[width=0.49 \textwidth]{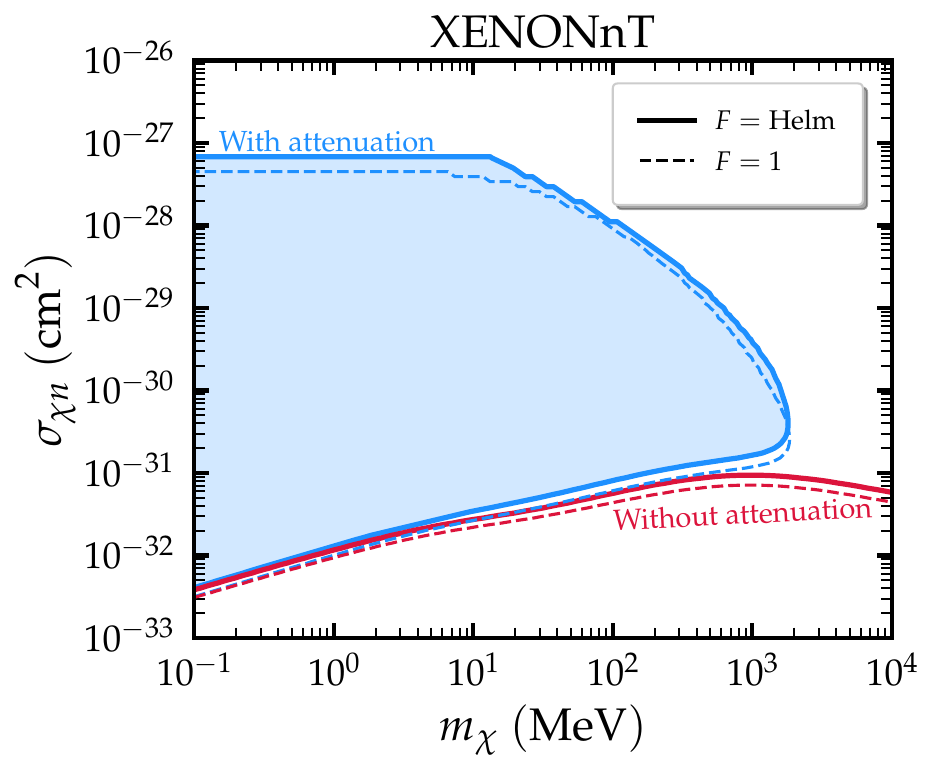}
  \includegraphics[width=0.49 \textwidth]{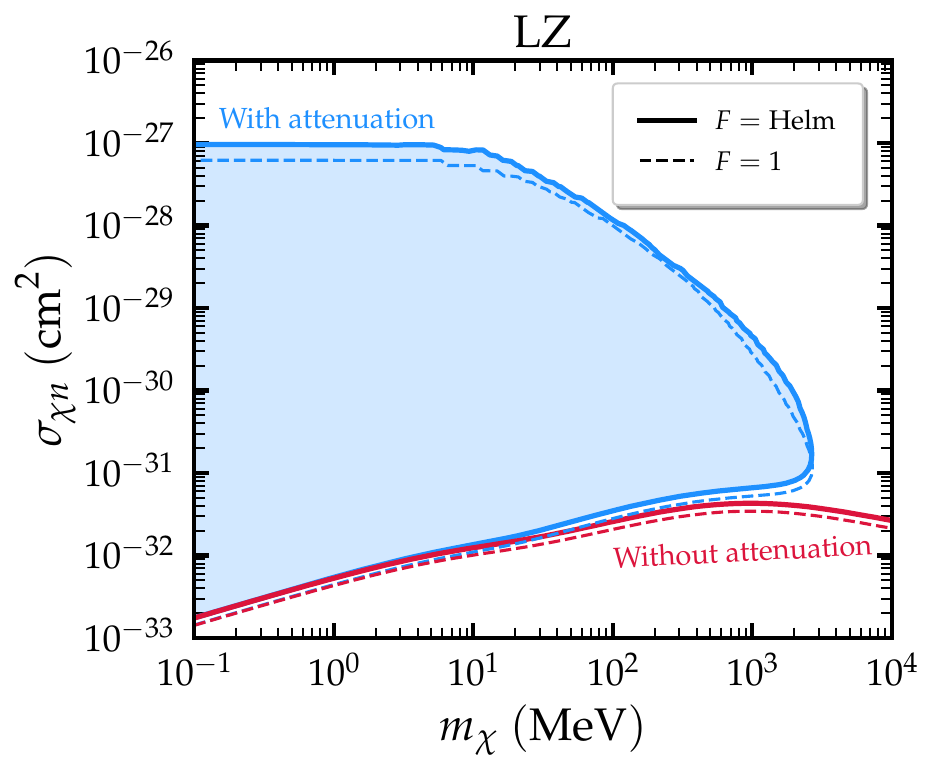}
  \caption{90\% C.L. exclusion limits on $\sigma_{\chi n}$ and $m_\chi$, obtained including (blue closed contour) or neglecting (red line) attenuation effects due to DM scattering off Earth's elements before reaching the detectors. Left panel shows bounds obtained with XENONnT data, while right panel refers to LZ data.  Additionally, the impact of the nuclear form factor is shown, with solid lines representing the Helm form factor, and dashed lines obtained assuming $F=1$ for both attenuated and unattenuated scenarios.} 
 \label{fig:DM_n_Comparison}
 \end{figure}

To understand the shape of our bounds in more detail, we now explore the implications of considering attenuation effects and adopting a realistic nuclear form factor. We examine how these factors influence the exclusion limits on the scattering cross section and DM mass.
The energy loss experienced by the DSNB-boosted DM due to its scattering with the Earth's material introduces a significant impact on the derived exclusion limits. Figure~\ref{fig:DM_e_Comparison} illustrates the consequences of considering (blue, closed contour) or neglecting (red exclusion line) attenuation effects for $\chi\mathrm{-}e$ scattering, while Fig.~\ref{fig:DM_n_Comparison} depicts the same for $\chi\mathrm{-}\mathcal{N}$ scattering. Clearly, attenuation effects impose an upper bound on the exclusion region. Above a certain scattering cross section, energy loss becomes substantial such that the DSNB-boosted DM particles cannot be detected because of severe attenuation. This fact confirms the necessity of properly accounting for Earth's attenuation effects to ensure accurate and robust constraints on the DNSB-boosted DM parameters. Finally, for low DM mass, e.g. $m_\chi \lesssim 1$~MeV, we find that the exclusion regions which incorporate attenuation effects (blue contours) are excluding lower cross sections (below $\sigma_{\chi e} = 10^{-31}~\mathrm{cm^2}$) compared to the case where attenuation effects are ignored (red contours) in Fig.~\ref{fig:DM_e_Comparison}. This is expected since in the low-mass region, for a fixed $m_\chi$ the number of events corresponding to the attenuated case is larger compared to the unattenuated one. The reason behind this behavior is twofold: First for low $m_\chi$ the DM flux reaching the detector is larger at low $T_\chi$ when attenuation effects are taken into account in comparison to the unattenuated case. Second, the non-vanishing attenuated DM flux at lower $T_\chi$ is triggering lower recoils $T_e$ in the detector, which due to their inversely proportional dependence on the differential cross section given in Eq.~(\ref{eq:DM-nucleus-electron}) are eventually leading to an enhancement of the event rates given by Eq.~(\ref{equn:Diff_Events_Rate}). The same reasoning applies for the case of nuclear recoils, discussed below, but due to kinematics this effect is not visible in Fig.~\ref{fig:DM_n_Comparison}. Further details are given in Appendix~\ref{appendix3}.

Furthermore, for the case of $\chi\mathrm{-}\mathcal{N}$ scattering, considering a realistic nuclear form factor [see e.g. Eq.~\eqref{equn:DM-N_SI_xSec}] introduces visible effects, as illustrated in Fig.~\ref{fig:DM_n_Comparison}.
The inclusion of the Helm form factor effectively modifies the energy-loss dynamics compared to the $F=1$ scenario. Indeed, the finite nuclear size reduces the differential DM-nucleus cross section given in Eq.~\eqref{eq:DM-nucleus-electron}, thus leading to  a decrease in the energy-loss rate $dT_\chi^z/dz$ when a larger momentum transfer is involved i.e. for the high-energy tail of DSNB-boosted DM flux (for an illustration see  Fig.~\ref{fig:Txz_nucleon} in Appendix~\ref{appendix2}). This modification results in a shift of the upper bound of the exclusion region, allowing for slightly higher $\sigma_{\chi n}$ values before energy loss renders particles undetectable.

Such an exploration of the interplay between attenuation effects and nuclear-physics considerations leads to a more comprehensive and robust understanding of the complex dynamics governing DSNB-boosted DM scatterings. These insights emphasize the significance of accounting for the latter effects in the accurate interpretation of experimental results, providing insights into the implications of $\chi\mathrm{-} \mathcal{N}$ scattering. 

Before closing our discussion we should stress that obtaining large, $\mathcal{O} (10^{-30})$ cm$^2$, DM-nucleon cross sections from conventional DM models is challenging, primarily due to the stringent bounds imposed on mediators coupled to nucleons, as pointed out in~\cite{Elor:2021swj, Bell:2023sdq}. A model-dependent study should be performed to thoroughly address the applicability of our limits. The interested reader is for instance referred to~\cite{Elor:2021swj}, where a sub-GeV DM candidate is presented, dubbed HYPER, that can accommodate large DM-nucleon cross sections. This example presents a  promising pathway to reconcile large direct detection cross sections with cosmological and laboratory observations.

\section{Conclusions}
\label{sec:conclusions}

In this work we have revisited the possibility that sub-GeV DM is boosted to semi-relativistic velocities through collisions with the DSNB. Such a very energetic component of the total DM flux, while subdominant, would be detectable at DM DD experiments thus amplifying their experimental reach. 

We have analyzed the most recent data from two cutting-edge DM experiments, LZ and XENONnT, and  we have obtained stringent constraints on the DSNB-boosted DM parameter space. For the first time, we have considered both electron and nuclear scatterings and obtained bounds on the relevant cross sections and DM mass. These new bounds extend the reach of typical DM DD searches to even lower DM mass ranges, and they are consistent with other searches for cosmic-ray boosted DM. In this regard, we have illustrated that due to the higher intensity of the DSNB flux in comparison to cosmic rays, the former allows to exclude a larger part of the available parameter space. We further point out that in obtaining our results for DM-nuclei scattering we reliably account for corrections due to the finite nuclear size by incorporating a Helm-type nuclear form factor.

Our results hence complement other existing searches for sub-GeV DM. Most of all, they show that even with their very first and limited exposure time data sets, the low-threshold XENONnT and LZ experiments dominate the terrestrial limits on DM-nucleus scattering at very low DM masses, with good complementarity to neutrino experiments like Super-Kamiokande and cosmological observations.
Finally we have highlighted the importance of including Earth's attenuation effects in the analysis. In particular, we have demonstrated that they have a strong impact on the upper bound of our derived exclusion regions disfavoring large DM scattering cross sections, namely $\sigma_{\chi e} \gtrsim 2 \times 10^{-28}~\mathrm{cm^2}$ and $\sigma_{\chi n} \gtrsim 8 \times 10^{-28}~\mathrm{cm^2}$ for $m_\chi = 0.1$~MeV. 

In summary our current analysis, by taking into account the Earth's attenuation and finite nuclear size effects,  provides accurate and robust constraints on the parameter space of low mass DSNB-boosted DM, using the first data sets of LZ and XENONnT.

\acknowledgments

We are grateful to Filippo Sala for useful comments on this manuscript.
We acknowledge the use of the high-performance computing facilities offered by the Bhaskara Cluster at IISER Bhopal, which significantly facilitated the completion of this research.
VDR acknowledges financial support by the CIDEXG/2022/20 grant (project ``D'AMAGAT'') funded by Generalitat Valenciana and by the Spanish grant PID2020-113775GB-I00 (MCIN/AEI/10.13039/ 501100011033).
AM is grateful for the invaluable financial support provided by the Prime Minister Research Fellowship (PMRF), sponsored by the Government of India (PMRF ID: 0401970).
The work of DKP was supported by the Hellenic Foundation for Research and Innovation (H.F.R.I.) under the “3rd Call for H.F.R.I. Research Projects to support Post-Doctoral Researchers” (Project Number: 7036).
The work of RS has been supported by the SERB, Government of India grant SRG/2020/002303.

\appendix 

\section{\label{appendix1} Geometry of an underground detector's location}

\begin{figure}[ht!]
  \includegraphics[width=0.45 \textwidth]{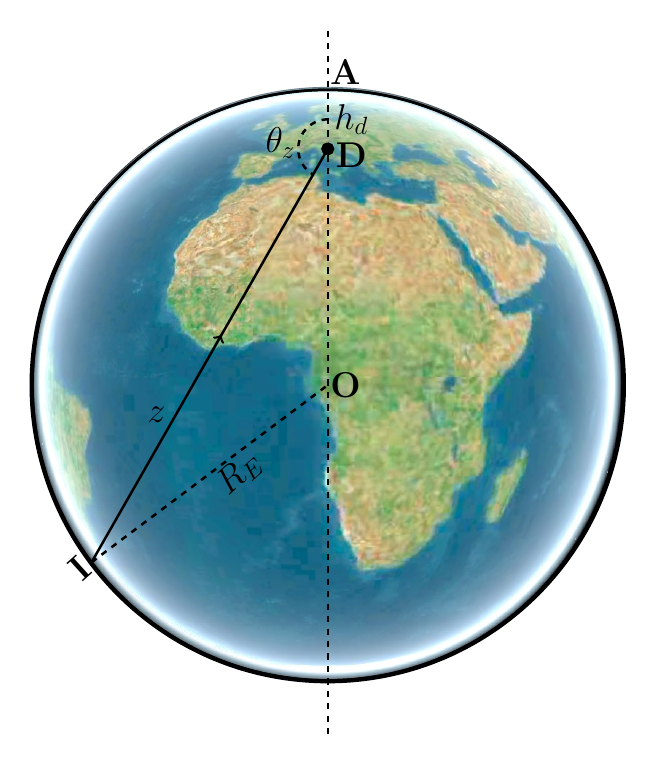}
  \caption{Geometry of the underground detector's location at a depth $h_d$ below Earth's surface, and  distance $z$ traveled by DM before detection.}
 \label{fig:zenith}
 \end{figure}
 In the context of DM detection, the geometry of the detector's underground position plays a pivotal role in assessing the anticipated signal rate and characteristics of the search. We explore the configuration of a detector located at a depth $h_d$ below the Earth's surface, as depicted in Fig.~\ref{fig:zenith}~\footnote{Credit of the underlying 3D world map: Apple Maps v 3.0.}. Our primary concern is to determine the  distance $z$ that corresponds to the distance traveled by DM particles between a specific impact point on the Earth's surface $(I)$ and the detector $(D)$. This distance depends on the zenith angle, $\theta_z$, representing the angle between the vertical direction and the line connecting the detector to the chosen point on the Earth's surface.

For the purpose of computing the distance $z$  we employ the law of cosines within the triangle $\triangle \mathrm{ODI}$ (see Fig.~\ref{fig:zenith}), leading to the following expression:

\begin{equation}
\begin{split}
R_E^2&=z^2+(R_E-h_d)^2-2R_E(R_E-h_d)\cos{(\pi-\theta_z)}\\
\Rightarrow z&=-(R_E-h_d)\cos{\theta_z}+\sqrt{R_E^2-(R_E-h_d)^2\sin^2{\theta_z}}\,.
\end{split}
\end{equation}
By means of this expression, we can precisely determine the distance $z$ that characterizes the spatial relationship between the position of the detector and the Earth's surface point of interest for a given zenith angle. Understanding this geometric configuration is crucial for a more realistic simulation of the interactions between DM particles and the detector and in predicting potential signals in DM experiments.

\section{\label{appendixB} Geophysical properties of Earth}

We model the Earth's interior as a sphere of constant electron and nuclear densities
($n_e=8\times 10^{23} \mathrm{~cm}^{-3}$ and $n_\mathcal{N} =  3.44 \times 10^{22} \mathrm{~cm}^{-3}$), based on the abundances of the main elements as shown in Table~\ref{Tab:Earth_elements}.

\begin{table}[ht]
    \centering
    \begin{tabular}{|@{\hspace{10pt}} c @{\hspace{20pt}} c @{\hspace{20pt}} c @{\hspace{20pt}} c@{\hspace{10pt}}|}
        \hline
        \textbf{Element} & \textbf{Mass Number ($\mathbf{A}$)} & \textbf{Relative Abundance (\%)} & $\mathbf{n_\mathcal{N}~\left(\mathrm{\textbf{cm}}^{-3}\right)}$ \\[0.1cm]
        \hline
        Fe & $55.845$ & $32.1$ & $6.11 \times 10^{22}$ \\
        O & $15.999$ & $30.1$ & $3.45 \times 10^{22}$ \\
        Si & $28.086$ & $15.1$ & $1.77 \times 10^{22}$ \\
        Mg & $24.305$ & $13.9$ & $1.17 \times 10^{22}$ \\
        S & $32.065$ & $2.9$ & $2.33 \times 10^{21}$ \\
        Ca & $40.078$ & $1.5$ & $7.94 \times 10^{20}$ \\
        Al & $26.982$ & $1.4$ & $1.09 \times 10^{21}$ \\
        \hline
    \end{tabular}
    \caption{Properties of the most abundant elements in the Earth's geosphere. The table presents the main elements found within the Earth's crust, mantle, and core~\cite{morgan, MCDONOUGH2003547}. For each element, the respective mass number ($A$), relative abundance, and nuclear number density ($n_\mathcal{N}$) are provided.}
            \label{Tab:Earth_elements}
\end{table}

\section{\label{appendix2}Energy loss experienced by the DSNB-boosted DM due to scattering inside Earth}

\begin{figure}[H]
  \includegraphics[width=0.49 \textwidth]{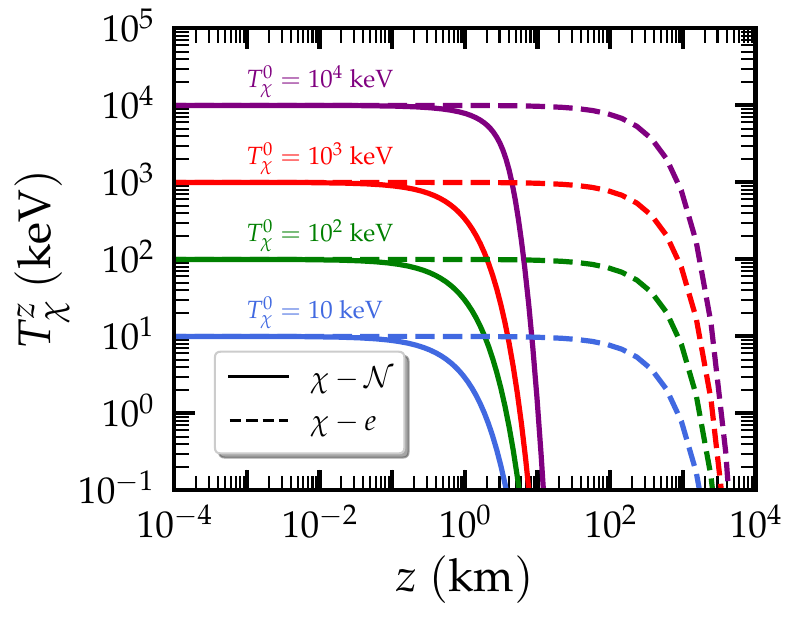}
  \includegraphics[width=0.48 \textwidth]{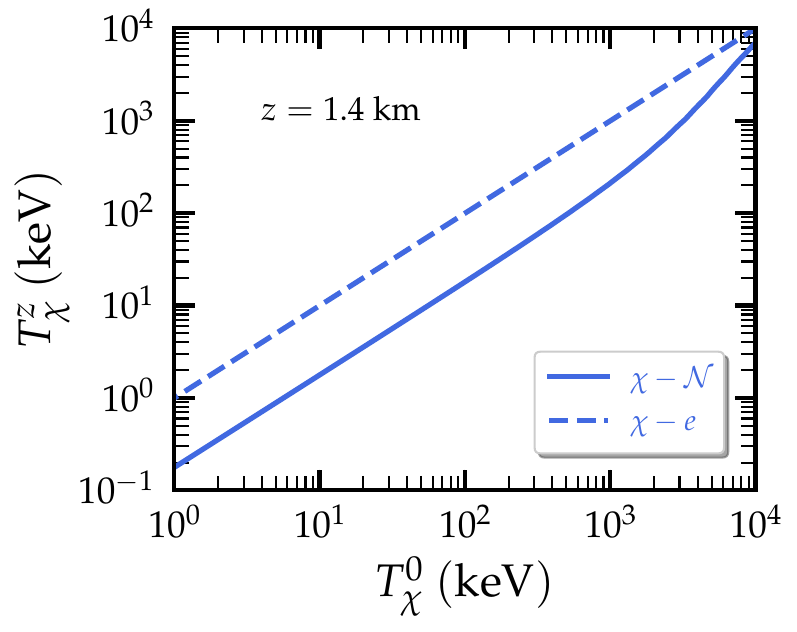}
  \caption{Left panel: $T_\chi^z$ as a function of $z$ for fixed $T_\chi^0$. Right panel: $T_\chi^z$ versus $T_\chi^0$ at fixed $z=1.4$ km. We fix $m_\chi = 300\mathrm{~MeV}$ and $\sigma_{\chi e} (\sigma_{\chi n}) = 10^{-29}\mathrm{~cm}^2$. Solid lines correspond to DM-nuclei scattering, while dashed lines represent DM-electron scattering.}
 \label{fig:Txz_single}
 \end{figure}

\begin{figure}[h!]
  \includegraphics[width=0.49 \textwidth]{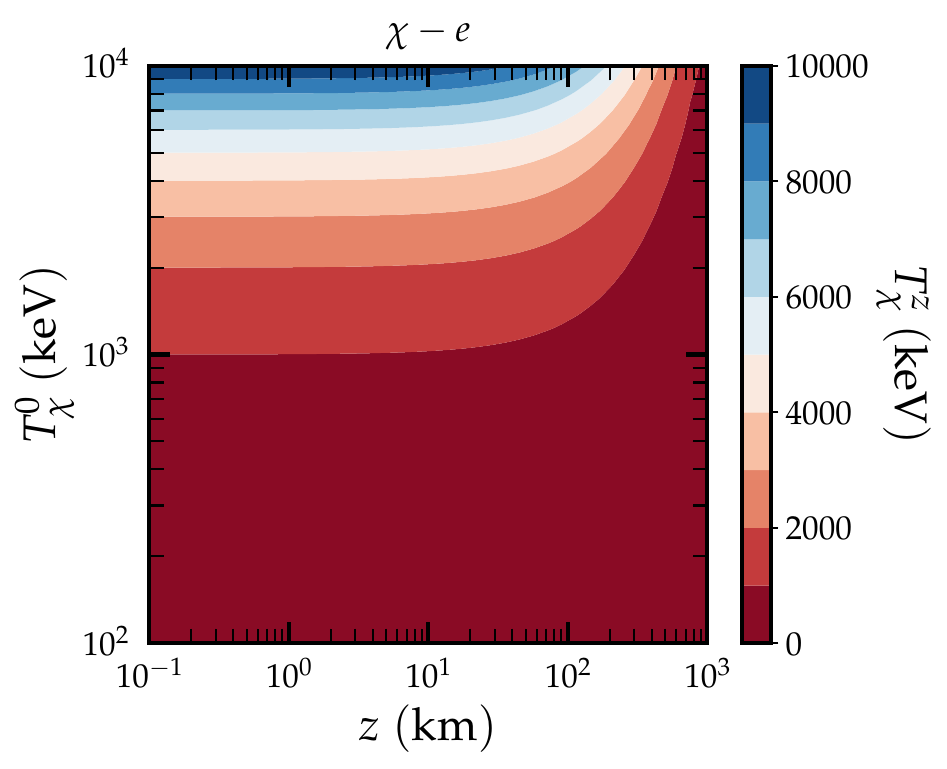}
  \includegraphics[width=0.49 \textwidth]{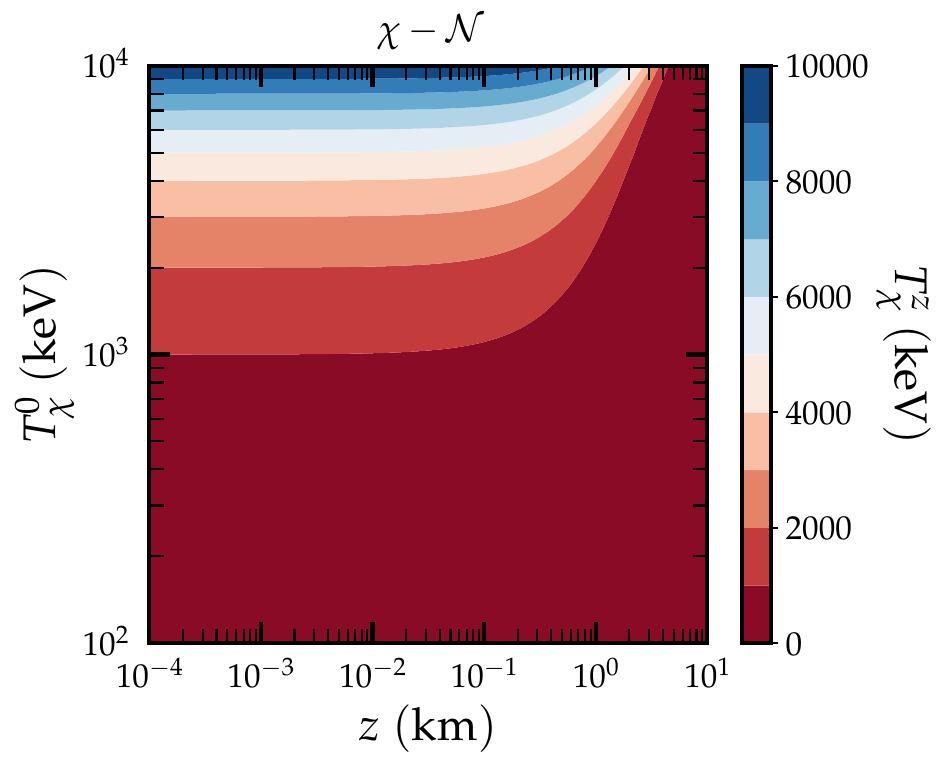}
  \caption{Contour plot of values of $T_\chi^z$ in the ($z$, $T_\chi^0$) plane, for $m_\chi = 300\mathrm{~MeV}$ and $\sigma_{\chi e} (\sigma_{\chi n}) = 10^{-29}\mathrm{~cm}^2$. The left panel refers to DM-electron scattering, while the right panel refers to DM-nuclei scattering. Both panels depict the variation of $T_\chi^z$ across different values of $T_\chi^0$ and $z$, providing insights into the interplay between these parameters.}
 \label{fig:Txz_double}
 \end{figure}

In this Appendix we investigate the dependence of the DSNB-boosted DM's underground kinetic energy (here denoted as $T_\chi^z$) on the distance ($z$) and initial value of the DM kinetic energy at Earth's surface ($T_\chi^0$). Such considerations provide essential insights into the impact of attenuation effects within Earth's materials. The left panel of Fig.~\ref{fig:Txz_single} illustrates $T_\chi^z$ as a function of $z$, derived by solving Eq.~\eqref{equn:Attenuation}, while considering different initial values of $T^0_\chi$. This analysis is performed separately for $\chi-e$ (dashed curves) and $\chi-\mathcal{N}$ (solid curves) scattering scenarios. As can be seen, the energy of the DSNB-boosted DM undergoes a rapid decrease for distances larger than $\sim 1\mathrm{~km}$ for $\chi-\mathcal{N}$ scattering, while in the case of $\chi-e$ scattering, this substantial energy reduction takes place at distances beyond $\sim100\mathrm{~km}$. In the right panel of Fig.~\ref{fig:Txz_single}, we present $T_\chi^z$ as a function of the initial kinetic energy $T^0_\chi$, computed at a fixed depth $z=1.4$~km (typical of DD experiments like XENONnT and LZ), i.e. corresponding to the special case $\theta_z=0$ for which $z=h_d$. This plot offers a direct comparison of the energy evolution assuming different initial conditions. The intricate interplay between these parameters is further visualized in the contour plot displayed in Fig.~\ref{fig:Txz_double}, which depicts the variation of $T_\chi^z$ across the $(z,T_\chi^0)$ plane. In all cases, we adopt a representative DSNB-boosted DM particle mass  $m_\chi = 300\mathrm{~ MeV}$ and assume a cross section $\sigma_{\chi e} = \sigma_{\chi n} = 10^{-29}\mathrm{~cm}^2$.

\begin{figure}[ht!]
  \includegraphics[width=0.49 \textwidth]{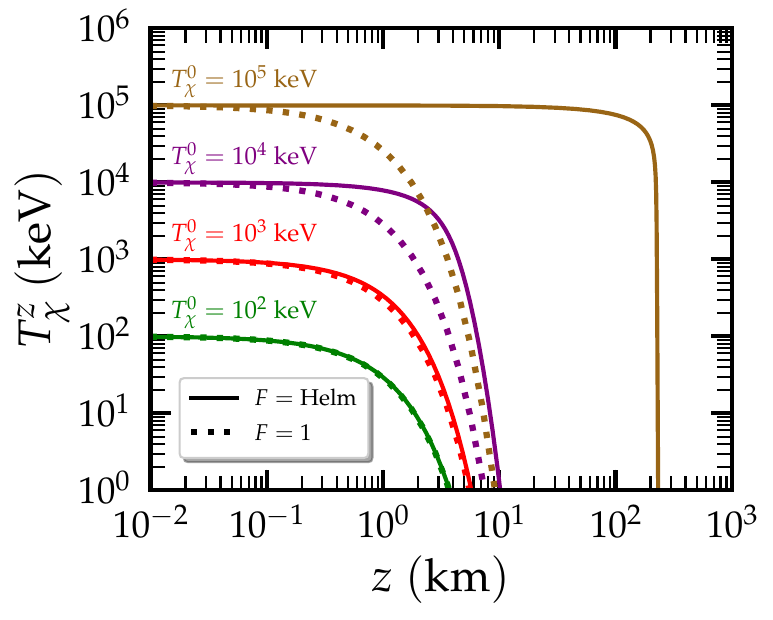}
  \caption{$T_\chi^z$ as a function of the distance $z$, for different initial kinetic energies of DM particles at  Earth's surface ($T_\chi^0$), focusing on $\chi-\mathcal{N}$ scattering interactions. The DM particle mass and cross section are fixed to $m_\chi= 300\mathrm{~MeV}$ and $\sigma_{\chi n}=10^{-29}\mathrm{~cm}^2$. The depicted results are obtained considering both the Helm form factor (solid lines) and $F=1$ (dotted lines).}
 \label{fig:Txz_nucleon}
 \end{figure}

 For the case of $\chi-\mathcal{N}$ scattering, incorporating finite nuclear size effects through the  nuclear form factor becomes particularly relevant. Figure~\ref{fig:Txz_nucleon} shows the impact of nuclear effects by considering two distinct scenarios in the calculation of the final kinetic energy of DM particles reaching the detector: (i) including a Helm-type nuclear form factor  and (ii) completely neglecting nuclear effects, i.e. $F=1$. Notably, the effect driven by nuclear physics becomes evident around $z=0.1\mathrm{~km}$. In particular for high-energy DSNB-boosted DM particles, the disparity between the two scenarios becomes substantial. All in all, incorporating nuclear physics effects through the Helm form factor leads to a reduction of the total cross section, resulting in a mitigated energy loss. Remarkably, DSNB-boosted DM particles with kinetic energies exceeding $100\mathrm{~MeV}$ undergo such a marginal energy loss that their energy remains nearly constant at distances around $z\lesssim 100~\mathrm{km}$.

\section{\label{appendix3}Behavior of the exclusion regions in the low DM mass regime}
Here we provide further clarifications regarding the behavior of the exclusion regions observed in Fig.~\ref{fig:DM_e_Comparison}, where the sensitivity of XENONnT and LZ experiments to DM-electron scattering was presented assuming both attenuated and unattenuated DSNB-boosted DM fluxes.
For relatively low DM masses $m_\chi$, the  exclusion limit in Fig.~\ref{fig:DM_e_Comparison} corresponding to DM-electron scattering when attenuation effects are taken into account (blue contours)  is extending to lower cross section values compared to the unattenuated  case (red contours). As we will explain, this is a direct consequence related to the shape of the energy distribution of DSNB-boosted DM particles reaching the terrestrial detector as well as to the recoil dependence of the differential event rates.

 For low values of $T_\chi$ the attenuated flux becomes significantly larger in comparison to the unattenuated one, a behaviour that is more pronounced when the value of $m_\chi$ gets smaller. Moreover, a non-vanishing DM flux at lower values of $T_\chi$ results into detectable signals of lower recoil energies $T_e$. As explained in the main text, the number of detected events is inversely proportional to $T_e$, see e.g. Eqs.~(\ref{eq:DM-nucleus-electron}, \ref{equn:Diff_Events_Rate}), thus a signal enhancement is expected for lower recoils. As a consequence, the events originating from the attenuated flux are enhanced in comparison to those coming from the unattenuated flux, affecting  the sensitivity in the parameter space of $(m_\chi, 
 \sigma_{\chi e})$ accordingly. This is particularly relevant for small values of $m_\chi$ (especially for $m_\chi \lesssim 1$ MeV). The opposite behavior occurs for larger values of $m_\chi$ (see the discussion below).
 
\begin{figure}[ht!]
  \includegraphics[width=0.49 \textwidth]{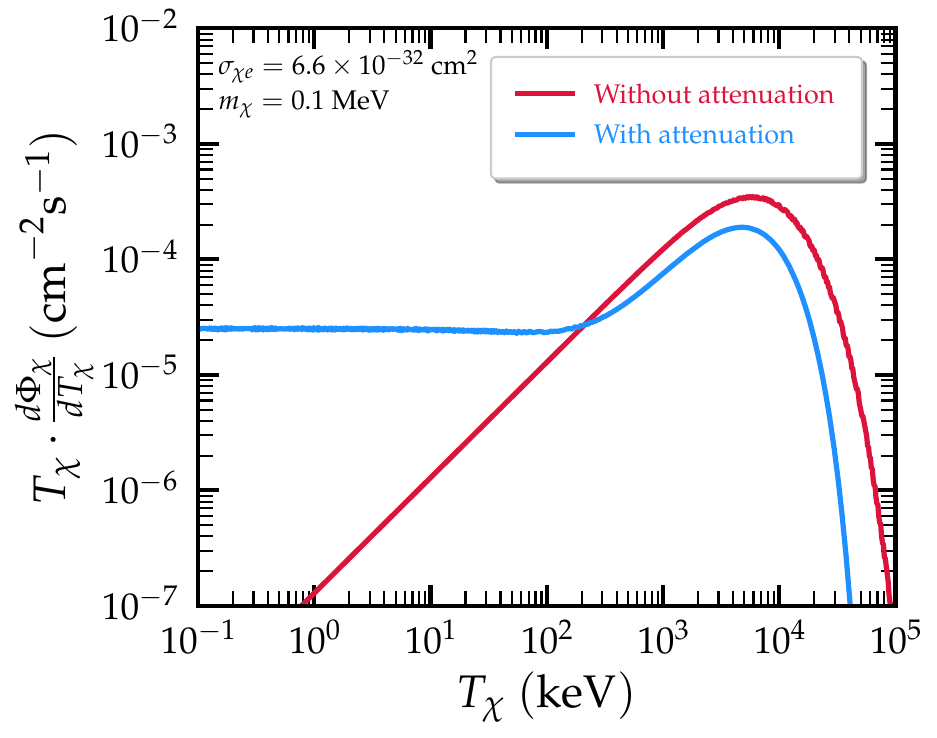}
  \includegraphics[width=0.44 \textwidth]{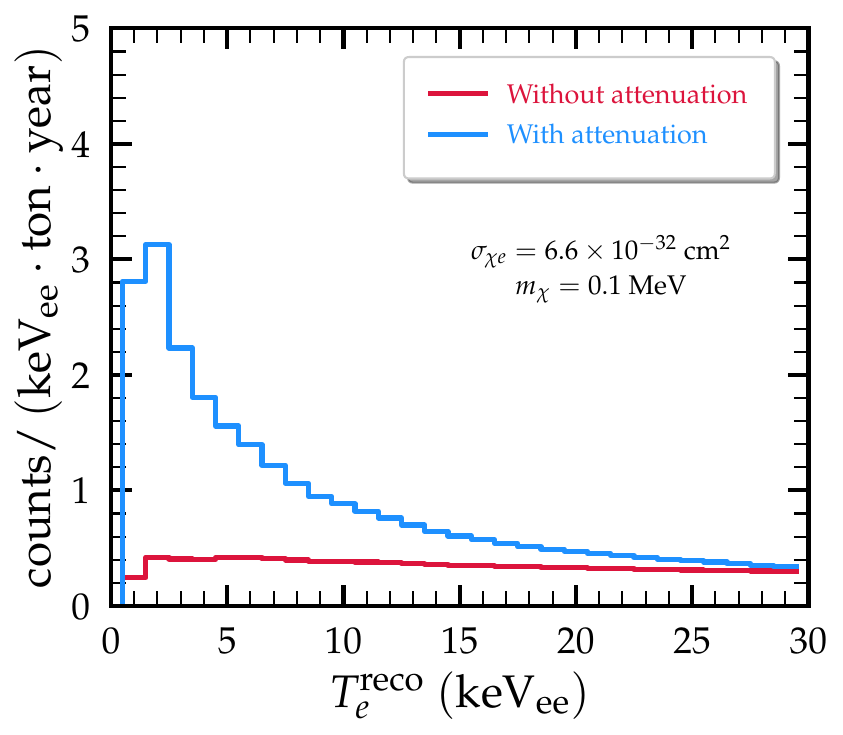}
    \includegraphics[width=0.49 \textwidth]{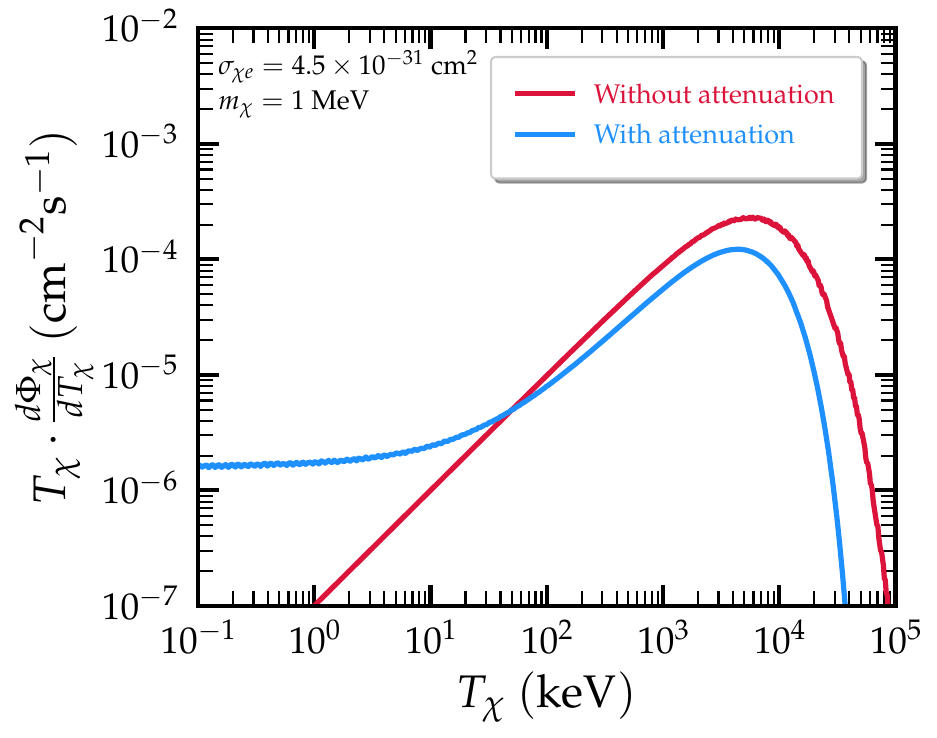}
  \includegraphics[width=0.44 \textwidth]{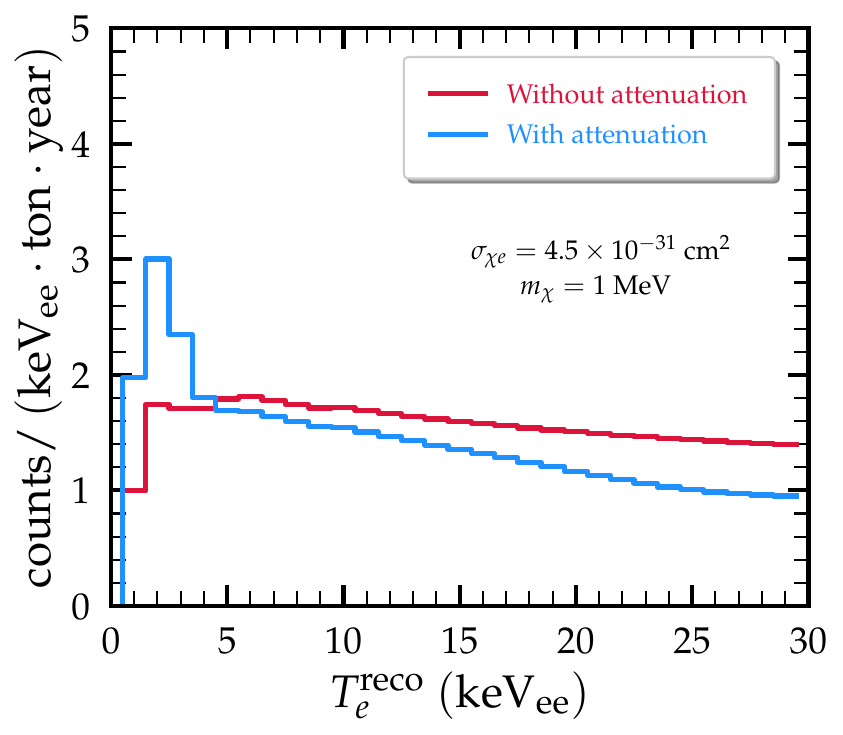}
   \includegraphics[width=0.49 \textwidth]{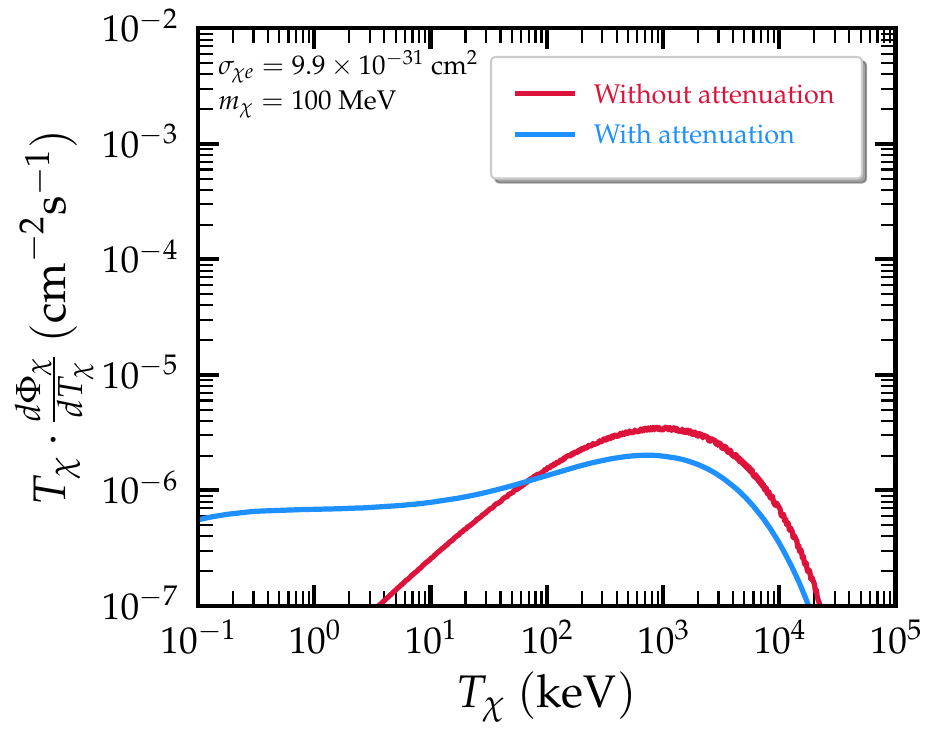}
  \includegraphics[width=0.44\textwidth]{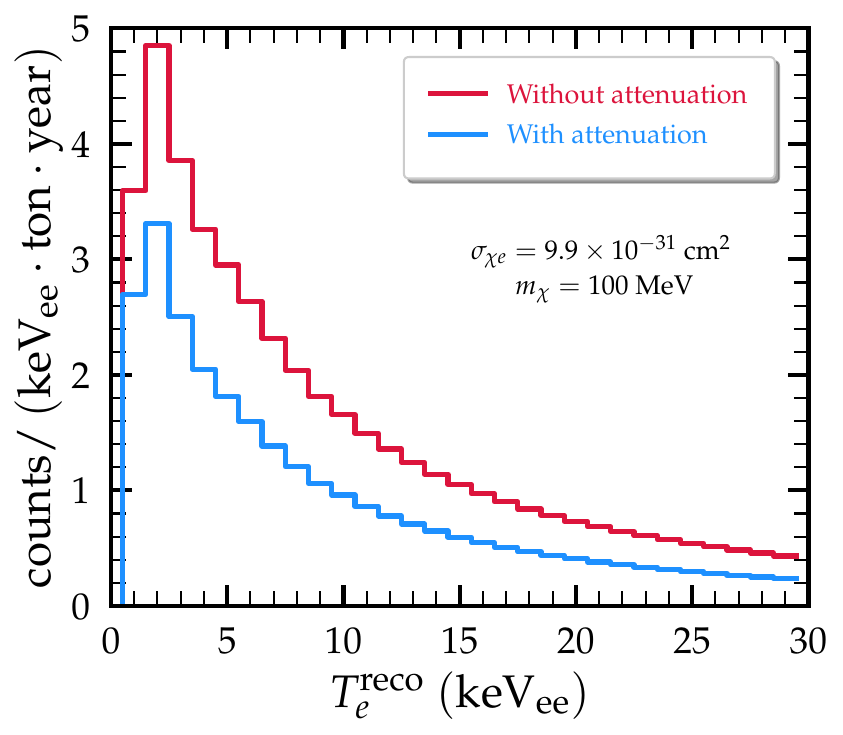}
  \caption{Left: Differential DSNB-boosted DM flux distribution as a function of the DM energy ($T_\chi$) for both attenuated and unattenuated scenarios. Right: Simulated differential events for the XENONnT experiment in response to DSNB-boosted DM interactions with electrons, corresponding to both attenuated and unattenuated scenarios. From top to bottom rows the DM mass is fixed at $m_\chi=0.1,~1$ and $~100\mathrm{~MeV}$, while the corresponding cross section values are obeying the exclusion limits of Fig.~\ref{fig:DM_e_Comparison} and read  $\sigma_{\chi e}=6.6\times 10^{-32},~4.5\times 10^{-31}$ and $9.9\times 10^{-31}~\mathrm{cm}^2$, respectively.}
 \label{fig:Flux_Events__Appendix}
 \end{figure}

A detailed illustration of these effects is presented in Fig. \ref{fig:Flux_Events__Appendix}. Each row from top to bottom corresponds to different values of DM mass $m_\chi = 0.1, 1, 100$~MeV, while left and right panels show the differential DSNB-boosted DM flux distribution and the respective differential event rates at XENONnT \footnote{For the case of LZ the results behave similarly and are omitted from Fig.~\ref{fig:Flux_Events__Appendix} due to space restriction.}. For each case, a comparison is given between the attenuated and unattenuated cases, while the chosen cross section values are allowed by the exclusion limits in Fig.~\ref{fig:DM_e_Comparison}. Results in the left panels of Fig. \ref{fig:Flux_Events__Appendix}  demonstrate that the lower the DM mass the larger the attenuated flux becomes in the low $T_\chi$ regime. In the right panels, the corresponding event rates show that the lower the DM mass the larger the enhancement of the events coming from the attenuated flux becomes compared to the unattenuated case. Let us finally note that for $m_\chi=100~\mathrm{MeV}$, the unattenuated flux is larger and so is the respective number of events;  consequently the exclusion region that corresponds to the unattenuated flux is slightly more constraining at this value of $m_\chi$. 

Before closing, let us stress that the obtained results highlight the intricate nature of direct detection of sub-GeV DM and the importance of considering these factors when interpreting experimental results.

\newpage
\section{\label{appendixE} Overview of the gauged $U(1)_{B_1-3L_e}$ extension}

In Sec.~\ref{sec:results}, we have presented our results for the benchmark case $\sigma_{\nu\chi}=\sigma_{\chi e}$ or $\sigma_{\nu\chi}=\sigma_{\chi n}$. This specific choice allows us to derive constraints on DM scattering in a simplified, yet representative manner. However, as explained in Sec.~\ref{sec:results}, our results can be easily recast into scenarios beyond this benchmark.

The critical point to be addressed is whether this benchmark choice can be realized in Ultra-Violet (UV) complete models for DM. As we will explain below, the answer is affirmative. This benchmark choice can be realized within the SM by introducing  additional scalar and vector leptons and/or quarks and writing down Yukawa interactions for SM fermions and DM. However, for such scalar-mediated scenarios, our benchmark case can be obtained only by requiring a fine-tuning of the Yukawa couplings of the different fermions, which might not be theoretically appealing. 

In contrast, extending the SM with a new gauge symmetry and assigning different charges to the fermions provides a more elegant solution. This approach naturally leads to comparable cross sections for DM scattering off different fermions, just by choosing appropriate charges for the fermions, without the need for any fine-tuning. Therefore, while the SM extension with an additional scalar provides a valid framework, the introduction of a new gauge symmetry offers a more compelling and theoretically appealing scenario.

For the reader's convenience in this Appendix we provide details regarding the gauged $U(1)_{B_1-3L_e}$ SM extension (motivated by Ref.~\cite{Bonilla:2017lsq}), which constitutes a framework that can accommodate interactions where the DM scattering cross sections with neutrinos, electrons, and nucleons are comparable. The choice of this particular flavor-dependent gauge extension of the SM is motivated by the desire to achieve comparable DM scattering cross sections with neutrinos, electrons, and nucleons. In DM-electron and DM-nucleon scattering, only one generation of leptons and quarks is involved, respectively. However, in DM-neutrino scattering, all three generations of neutrinos can contribute. Therefore, to ensure comparable cross sections, it becomes necessary to assign charges generation by generation, a feature naturally accommodated in our chosen model. Details are provided below.

The charges of SM quark ($Q_i$) and lepton ($L_i$) doublets as well as the quark and lepton singlets ($u_i, d_i, l_i$) and right handed neutrinos $\nu_{i_R}$, where $i =1,2,3$ denotes the generation index, are given in Table~\ref{tab:B_1-3L_e_Charges}.

\begin{table}[ht]
\centering
\begin{tabular}{|@{\hspace{10pt}} c @{\hspace{50pt}} c @{\hspace{50pt}} c@{\hspace{10pt}}|}
\hline
\textbf{Fields} & $\mathbf{SU(3)_c \otimes SU(2)_L \otimes U(1)_Y}$ & $\mathbf{U(1)_{B_1-3L_e}}$ \\
\hline
$L_1 \equiv L_e$ & ($1,2,-1$) & $-3$ \\
$l_1 \equiv e_R$ & ($1,1,-2$) & $-3$ \\
$\nu_{e_R}$ & ($1,1,0$) & $-3$ \\
$L_{\mu/\tau}$ & ($1,2,-1$) & $0$ \\
$\mu_R/\tau_R$ & ($1,1,-2$) & $0$ \\
$\nu_{\mu_R/\tau_R}$ & ($1,1,0$) & $0$ \\[0.1cm]
\hline
$Q_{1}$ & ($3,2,\frac{1}{3}$) & $1$ \\
$u_{1}$ & ($3,1,\frac{4}{3}$) & $1$ \\
$d_{1}$ & ($3,1,-\frac{2}{3}$) & $1$ \\
$Q_{2/3}$ & ($3,2,\frac{1}{3}$) & $0$ \\
$u_{2/3}$ & ($3,1,\frac{4}{3}$) & $0$ \\
$d_{2/3}$ & ($3,1,-\frac{2}{3}$) & $0$ \\[0.1cm]
\hline
$H$ & ($1,2,1$) & $0$ \\
$\phi$ & ($1,1,0$) & $y$ \\
\hline
$\chi$ & ($1,1,0$) & $x$ \\
\hline
\end{tabular}
\caption{Matter content and charge assignment of the gauged $B_1-3L_e$ model.}
\label{tab:B_1-3L_e_Charges}
\end{table}

In Table~\ref{tab:B_1-3L_e_Charges}, in addition to SM fermions, three right-handed neutrinos have been added ($\nu_{e_R}$,  $\nu_{\mu_R}$, $\nu_{\tau_R}$), the inclusion of $\nu_{e_R}$ being required by anomaly cancellation conditions while the other two are introduced to facilitate neutrino mass generation. Regarding the scalar fields, apart from the SM Higgs doublet ($H$), the singlet field ($\phi$) is added for the Spontaneous Symmetry Breaking (SSB) of  $B_1-3L_e$ gauge symmetry, while the singlet scalar $\chi$ serves as the DM candidate. Unlike fermion charges, the scalar charges cannot be fixed by the anomaly cancellation conditions, and thus they are left free. However, their charges can be constrained by additional requirements e.g. DM stability as we will discuss below.

The $SU(2)_L \otimes U(1)_Y \otimes U(1)_{B_1-3L_e}$ invariant Lagrangian of the model is given as follows:

\begin{itemize}
    \item[$\blacksquare$] \textbf{Fermion-Gauge Sector:} The covariant derivative is defined as
    \begin{equation}
    \label{eq.covariant_derivative}
        D^\mu=\partial^\mu+igT^a W_a^\mu+i\frac{g'}{2}YB^\mu+ig_x Q_{B_1-3L_e}Z'^\mu\,,
    \end{equation}
    where $g_x$ is the $U(1)_{B_1-3L_e}$ gauge coupling strength, $Q_{B_1-3L_e}$ are the $B_1-3L_e$ charges of the fields listed in Table~\ref{tab:B_1-3L_e_Charges} and  $W^\mu_a, B^\mu, Z'^\mu$ are the weak, hypercharge, and $B_1-3L_e$ mediators respectively. The $W^\mu_3$ and $B^\mu$ fields will further mix to form the photon  $A^\mu$ and the $Z^\mu$. Hence the Lagrangian density in the Fermion-Gauge sector can be written as

    \begin{equation}
    \label{eq.Fermion_Gauge}
\mathscr{L}\subset \sum^3_{j=1} \left(i{\overline{Q}_{j}}\gamma_\mu D^\mu{Q_{j}}
+ i{\overline{u}_{j}}\gamma_\mu D^\mu{u_{j}} 
+ i{\overline{d}_{j}}\gamma_\mu D^\mu{d_{j}} 
+ i{\overline{L}_{j}}\gamma_\mu D^\mu{L_{j}} 
+ i{\overline{l}_{j}}\gamma_\mu D^\mu{l_{j}} 
+ i{\overline{\nu}_{j_R}}\gamma_\mu D^\mu{\nu_{j_R}}\right)\,.
    \end{equation}

\item[$\blacksquare$] \textbf{Gauge Kinetic Sector:} The gauge kinetic term is
\begin{equation}
\label{eq.Gauge_Kinatic}
    \mathscr{L}\subset -\frac{1}{4} W^{\mu \nu}_a W^a_{\mu \nu} - \frac{1}{4} B^{\mu \nu} B_{\mu \nu} - \frac{1}{4} Z'^{\mu \nu}Z'_{\mu \nu}\,,
\end{equation}
where $W^a_{\mu \nu}, B_{\mu\nu},$ and $Z'_{\mu\nu}$ are the  field strength tensors corresponding to the gauge groups $SU(2)_L$, $U(1)_Y$, and $U(1)_{B_1-3L_e}$ respectively.

\item[$\blacksquare$] \textbf{Yukawa Sector:} The Lagrangian density in the Yukawa sector is given by
\begin{equation}
\label{yukawa_Lagrangian}
\begin{aligned}
 \mathscr{L}\subset &-Y_d{\overline{Q}}H{d} 
 - Y_u{\overline{Q}}\widetilde{H}{u} 
 - Y_{l}{\overline{L}}H{l} - Y_{\nu}{\overline{L}}\widetilde{H}{\nu}_{R} 
 -  M_R \,\overline{{\nu_e}_{_R}^c}{\nu_e}_{_R}\phi+\mathrm{~h.c.}\,.
\end{aligned}
\end{equation}
    In Eq.~\eqref{yukawa_Lagrangian} we have suppressed the generation indices to avoid cluttering. 
    Note that the SM gauge charges  allow also  for the term $\overline{{\nu}_{R}^c}{\nu}_{R}\chi$, however, the presence of such a term would lead to the decay of $\chi$. This term can be easily forbidden by the $U(1)_{B_1-3L_e}$ symmetry which implies $x\neq 0$ and $x \neq |6|$.
    
\item[$\blacksquare$] \textbf{Scalar Sector:} The Lagrangian density in the scalar sector can be expressed as follows
    \begin{equation}
    \label{Scalar_Lagrangian}
        \begin{aligned}
            \mathscr{L}\subset &(D_\mu H)^\dagger (D_\mu H)  + (D_\mu \phi)^\dagger (D_\mu \phi) + (D_\mu \chi)^\dagger (D_\mu \chi) - V(H, \phi, \chi)\,.
        \end{aligned}
    \end{equation}
    The potential term $V(H, \phi, \chi)$ in \eqref{Scalar_Lagrangian} is given by
    \begin{equation}
    \label{Potential}
        \begin{aligned}
            V(H, \phi, \chi)& = \mu_H^2 H^\dagger H +\frac{\lambda_H}{2}(H^\dagger H)^2+\mu_\phi^2(\phi^\dagger \phi)+\frac{\lambda_\phi}{2}(\phi^\dagger \phi)^2+\mu_\chi^2(\chi^\dagger \chi)+\frac{\lambda_\chi}{2}(\chi^\dagger \chi)^2\\
            &+\lambda_{H\phi}(H^\dagger H)(\phi^\dagger \phi)+\lambda_{H\chi}(H^\dagger H)(\chi^\dagger \chi)+\lambda_{\phi\chi}(\phi^\dagger \phi)(\chi^\dagger \chi)+\mathrm{~h.c.}\,.
        \end{aligned}
    \end{equation}
     DM decay terms such as $(H^\dagger H)\chi^\dagger \phi,~(H^\dagger H)\chi \phi,~(\phi^2)\chi,~(\phi^\dagger)^2\chi$ etc., can be easily avoided by choosing $x\neq |y|$, and $x\neq 2|y|$ in the potential, Eq.~\eqref{Potential}. Additionally, it is worth noting that while $H$ and $\phi$ acquire vacuum expectation values (vev) to spontaneously break $SU(2)_L \otimes U(1)_Y \otimes U(1)_{B_1-3L_e}$ symmetry, the DM candidate $\chi$ does not acquire a vev.
\end{itemize}

Thus, in summary the $U(1)_{B_1-3L_e}$ charge $x$ of the DM $\chi$ can take any value except $x = 0$, $x = |6|$, $x = |y|$ and $x = 2|y|$.

The cross sections for DM scattering with electrons, nucleons, and neutrinos through $Z'$ mediator can be expressed as follows:
\begin{itemize}
    \item[$\bullet$]\textbf{\textit{DM-electron scattering ($\mathbf{\mathit{\sigma_{\chi e}}}$):}} DM particles can scatter off electrons in direct-detection experiments through $Z'$ exchange; the DM-electron cross section can be expressed in terms of the coupling~\cite{Essig:2015cda}
    \begin{equation}
        \sigma_{\chi e}=\frac{36x^2\mu_{\chi e}^2g_x^4}{\pi m_{Z'}^4}\,,
    \end{equation}
where $\mu_{\chi e}$ denotes the DM-electron reduced mass and $m_{Z'}$  the mediator mass.
    \item[$\bullet$]\textbf{\textit{DM-nucleon scattering ($\mathbf{\mathit{\sigma_{\chi n}}}$):}} Similarly the DM-nucleon cross section at zero momentum transfer can be written as~\cite{Martinez:2015wrp}
    \begin{equation}
        \sigma_{\chi n}=\frac{36x^2\mu_{\chi n}^2g_x^4}{\pi m_{Z'}^4}\,,
    \end{equation}
    where $\mu_{\chi n}$ stands for the DM-nucleon reduced mass.

    \item[$\bullet$]\textbf{\textit{Neutrino-DM scattering ($\mathbf{\mathit{\sigma_{\nu\chi}}}$):}} Finally, the neutrino-DM cross section can be expressed as~\cite{Olivares-DelCampo:2017feq, Pandey:2018wvh}
    \begin{equation}
        \sigma_{\nu\chi}= \frac{36x^2E_\nu^2g_x^4}{\pi m_{Z'}^4}\,.
    \end{equation}

\end{itemize}

Thus, the $B_1-3L_e$ gauge symmetry simultaneously provides a framework for the DM matter extension of SM and naturally
leads to our benchmark scenario. 


\bibliographystyle{utphys}
\bibliography{bibliography}

\end{document}